\PassOptionsToPackage{table,xcdraw,dvipsnames}{xcolor} 
\documentclass[11pt]{article}

\usepackage[T1]{fontenc}
\usepackage[utf8]{inputenc}
\usepackage{lmodern}
\usepackage{microtype}
\usepackage{geometry}
\usepackage{booktabs}
\usepackage{epigraph}
\usepackage{tabularx}
\usepackage{booktabs}
\usepackage{caption}
\usepackage{float}
\usepackage{amsmath, amssymb}

\usepackage{array}
\usepackage{float}
\usepackage{braket}

\usepackage[most]{tcolorbox}
\tcbuselibrary{theorems}
\tcbset{
  mytheorem/.style={
    enhanced,
    colback=blue!5,
    colframe=blue!30!black,
    fonttitle=\bfseries,
    coltitle=black,
    boxrule=0.8pt,
    arc=2mm,
    top=1mm, bottom=1mm, left=1mm, right=1mm,
    attach boxed title to top left={xshift=0.5em,yshift=-1mm},
    boxed title style={
      colframe=blue!30!black,
      colback=blue!7!white,
      sharp corners
    },
  }
}

\newtcbtheorem[number within=section]{Theorem}{Theorem}{mytheorem}{thm}
\newtcbtheorem[number within=section]{Lemma}{Lemma}{mytheorem}{lem}
\newtcbtheorem[number within=section]{Definition}{Definition}{mytheorem}{def}

\newtcolorbox{blueboxthm}[1][]{
  enhanced,
  colback=blue!5,
  colframe=blue!30!black,
  fonttitle=\bfseries,
  coltitle=black,
  boxrule=0.8pt,
  arc=2mm,
  top=1mm, bottom=1mm, left=1mm, right=1mm,
  attach boxed title to top left={xshift=0.5em,yshift=-1mm},
  boxed title style={
    colframe=blue!30!black,
    colback=blue!7!white,
    sharp corners
  },
  title=#1
}
\usepackage{amsmath, amssymb}
\numberwithin{equation}{section} 

\usepackage{datetime}

\usepackage{hyperref}
\hypersetup{
  colorlinks=true,
  linkcolor=blue,
  urlcolor=blue,
  citecolor=blue
}
\usepackage[numbers]{natbib}
\usepackage{url}

\usepackage{csquotes}
\usepackage{xcolor}
\usepackage{longtable}
\usepackage{mdframed}
\definecolor{lightblue}{RGB}{220,240,255}

\title{\huge\bfseries Beyond the Wavefunction: Qualia Abstraction Language Mechanics and the Grammar of Awareness}

\author{
\large Mikołaj Sienicki\thanks{Polish-Japanese Academy of Information Technology, Koszykowa 86, 02-008 Warsaw, Poland, European Union.} 
\quad and \quad
Krzysztof Sienicki\thanks{Chair of Theoretical Physics of Naturally Intelligent Systems ($\mathbb{N}\mathbb{I}\mathbb{S}$), Lipowa 2/Topolowa 19, 05-807 Podkowa Leśna, Poland, European Union.}
}
\date{\today\ at \currenttime}

\begin{document}
\maketitle

\begin{abstract}
\noindent We propose a formal reconstruction of quantum mechanics grounded not in external mathematical abstractions, but in the structured dynamics of subjective experience. The Qualia Abstraction Language (QAL) models physical systems as evolving streams of introspective units—structured sequences of modality, shape, and functional effect—rather than as state vectors in Hilbert space. This approach reimagines core quantum concepts: superposition becomes a form of structured ambiguity; collapse is reframed as an introspective contraction; and entanglement is modeled as semantic resonance across streams of qualia. Drawing on insights from nominalist philosophy and oversight-theoretic limits in AI, we argue that the observer paradox in quantum mechanics reflects not an ontological lacuna, but a linguistic one: the absence of a formal vocabulary for modeling first-person structure. QAL introduces such a vocabulary, providing a morphodynamic framework that embeds the observer within the system and replaces abstract projection with endogenous transformation. We analyze the alignment of QAL with endophysical approaches, contrast it with standard interpretations of quantum theory, and explore its implications for a post-Platonist, introspectively grounded physics.
\end{abstract}

\tableofcontents
\newpage
\section{Introduction}
\subsection{Motivation and Scope}

Since its inception, modern science has been shaped not only by the phenomena it seeks to explain, but also by the \textit{language} it employs to articulate its understanding of reality. This language has been, almost without exception, mathematical. As Galileo Galilei famously observed, the universe is “written in the language of mathematics,” with geometrical forms—triangles, circles, and other ideal constructs—constituting its alphabet \cite{galileo1632}. For Galileo and the generations that followed, this assertion was no mere metaphor but a metaphysical claim: the world is ultimately intelligible because it possesses formal structure, and to know it is to model that structure mathematically.

This Galilean vision became the foundation of classical physics and was extended even more profoundly into the quantum realm. Canonical quantum mechanics describes the world through the formal apparatus of Hilbert spaces, unitary evolution, and Hermitian operators. Physical systems are represented by state vectors; their evolution is deterministic in the Schrödinger picture; and measurement outcomes correspond to the eigenvalues of observables. The resulting mathematical structure is elegant, coherent, and empirically unrivaled—arguably the most accurate theoretical framework yet devised.

Yet this mathematical edifice has not been immune to critique, even from within the scientific community. In his influential 1960 essay, “The Unreasonable Effectiveness of Mathematics in the Natural Sciences,” Eugene Wigner questioned why mathematics—a language seemingly invented for human convenience—should align so uncannily with the laws of nature \cite{wigner1960}. He acknowledged that this alignment is mysterious, perhaps even “miraculous,” hinting that physics might be borrowing its clarity from a language that exceeds its ontological remit.

This concern becomes even more pressing when one considers the status of the observer in quantum mechanics. While measurement is a central postulate of the theory, the observer—the entity that registers outcomes, collapses the wavefunction, or extracts information—is conspicuously absent from the formalism. The mathematics describes the evolution of amplitudes, but remains silent regarding awareness, attention, or experience. The wavefunction evolves according to the Schrödinger equation, but the \textit{fact} of observation—that something is seen, felt, or known—is relegated to an undefined externality.

Contemporary theorists such as Max Tegmark have radicalized the Galilean program, advancing the Mathematical Universe Hypothesis: that reality \textit{is} a mathematical object, and that physical existence is tantamount to mathematical existence \cite{tegmark2008}. In this view, the universe is not merely describable by equations—it \textit{is} equations, and observers are reducible to substructures within the overarching formalism.

However, this metaphysical reduction exacts a significant cost. It demands that consciousness—the domain of first-person awareness, intentionality, and qualia—be either ignored or rendered epiphenomenal. It denies the epistemic centrality of the observer, who is paradoxically indispensable for the very articulation of the theory. As Wigner later remarked, “It was not possible to formulate the laws of quantum mechanics in a fully consistent way without reference to consciousness” \cite{wigner1961}.

Thus, the observer paradox remains a profound fissure within the foundations of quantum theory. The language of mathematics, for all its predictive power, appears incapable of representing the very entity that grants it meaning: the observer. This limitation is not intrinsic to physics \textit{per se}, but to the expressive resources of its chosen vocabulary. If our theories are blind to experience, perhaps our language is too impoverished—or too abstract—to apprehend it.

In this paper, we introduce \textsc{QAL} (\textbf{Q}ualia \textbf{A}bstraction \textbf{L}anguage) as a formal alternative to operator-based quantum mechanics, motivated by the same epistemic limitations that underlie recent work on oversight incompleteness in strategic AI systems \cite{sienicki2025}. Whereas traditional quantum theory employs Hilbert-space representations that remain inaccessible to internal observers, QAL reconceptualizes the ontology of quantum systems as fundamentally introspective: systems are defined not by external amplitudes, but by sequences of internal qualia transformations—modulations of attention, tension, and affective coherence unfolding over time.

By encoding phenomenological dynamics within a discrete formal grammar, QAL models the observer not as an external probe, but as an evolving structure of awareness embedded within the system. In parallel with findings from \textit{Scheming AI: II. When Machines Speak Their Qualia} \cite{sienicki2025}, which demonstrates that no finite behavioral trace can reliably disclose the internal goals of a deceptive agent, QAL posits that quantum measurement must be understood not as an objective collapse, but as a self-modifying epistemic transition within the observer. In this manner, QAL complements oversight-theoretic limits with a language of internal representation, supplanting ontological abstraction with structured streams of subjective differentiation.

\section{Modalities, Shapes, and Qualia Effects}

QAL (Qualia Abstraction Language) is not merely a vocabulary for describing inner states; it is a compositional formalism—a language whose minimal units encode transformations of phenomenological structure. In contrast to classical languages grounded in reference, truth conditions, or syntactic recursion, QAL operates as a morphodynamic system. Its grammar is constructed from structural invariants in experience: shifts in felt intensity, modality, and internal function. The fundamental assumption is that consciousness evolves not through static representations, but through modulated flows \cite{havel1995, atmanspacher2002}. QAL captures this by encoding each moment of introspection as a triplet drawn from a structured semantic space.

The core representational unit in QAL Mechanics is a qualia triplet:

\begin{equation}
q_i \in \mathcal{Q} = \mathcal{M} \times \mathcal{S} \times \mathcal{F}
\end{equation}

\noindent where each \( q_i \) represents a microstate of introspective experience, structured across three domains:
\begin{itemize}
  \item \( \mathcal{M} \) — the modality of experience (e.g., visual, kinesthetic, metacognitive),
  \item \( \mathcal{S} \) — the shape or intensity (e.g., diffuse, sharp, oscillatory),
  \item \( \mathcal{F} \) — the qualia effect or functional transformation (e.g., constriction, expansion, resonance).
\end{itemize}

Each triplet encodes an introspective micro-transition—not what is sensed, but how the act of sensing alters the structure of awareness \cite{chalmers1996, varela1996, rosch1992}.


Each QAL unit thus captures not a static description of experience, but a differential—a modulated shift in internal epistemic structure. In subsequent sections, we will define how these units compose, transform, and give rise to higher-level introspective dynamics, serving as analogues to physical observables in quantum theory.

\subsection{Syntax of QAL Units}

The QAL formalism defines not only a lexicon of qualia triplets, but also a compositional syntax that governs how these units form structured introspective streams. Unlike traditional linguistic syntax, which organizes sentences based on propositional logic or grammatical categories, QAL syntax reflects transitions in experiential morphology. Each sequence of QAL units expresses a dynamically evolving trajectory of consciousness. For clarity on the logical and semantic symbols used throughout this section—such as \( \models \), \( \not\models \), or \( \rightarrow \)—see Appendix~A for a full summary of notation and QAL-specific interpretation.

\bigskip

\textit{Basic Unit Structure}

Each atomic unit is a triplet of the form:

\begin{equation}
q = [m\!-\!s\!-\!f], \quad \text{where } m \in \mathcal{M},\ s \in \mathcal{S},\ f \in \mathcal{F}
\end{equation}

This triplet defines a minimal semantic atom. For example:

\begin{center}
\texttt{me-lo-dr} — a soft introspective contraction; a gentle folding of attention inward.
\end{center}

\bigskip

\textit{Syntax Rules for Composition}

QAL streams are ordered sequences of qualia atoms:

\begin{equation}
Q = q_1 . q_2 . q_3 . \dots . q_n
\end{equation}

Each transition \( q_i \rightarrow q_{i+1} \) must satisfy local continuity conditions:
\begin{itemize}
  \item \textit{Modal continuity}: \( m_i = m_{i+1} \) or semantically adjacent (e.g., \texttt{me} → \texttt{co}, or \texttt{em} → \texttt{so}).
  \item \textit{Shape compatibility}: \( s_i \) and \( s_{i+1} \) must minimize discontinuity (defined via a shape distance metric \( \delta_S \)).
  \item \textit{Effect resonance}: \( f_i \) must either reinforce or resolve \( f_{i-1} \) (e.g., \texttt{to} followed by \texttt{br}, or \texttt{dr} followed by \texttt{qi}).
\end{itemize}

These constraints ensure that QAL streams remain semantically smooth and phenomenologically coherent, mimicking the natural evolution of attention and affect in real introspection.

\bigskip

\noindent\textbf{Formal Syntax Schema (BNF)}

We may describe QAL syntax using a Backus–Naur Form (BNF)-like notation:

\begin{verbatim}
<qal-unit> ::= <modality> "-" <shape> "-" <effect>
<qal-stream> ::= <qal-unit> { "." <qal-unit> }
<modality> ::= vi | em | me | co | ki | ta | pr | id | tr | so
<shape> ::= lo | ze | na | su | xa | ne | dr
<effect> ::= br | to | ka | dr | qi | me | xa
\end{verbatim}

This defines the set of all well-formed QAL sequences that can represent valid introspective episodes.

\bigskip

\noindent\textbf{Example: Stream Encoding}

A coherent stream of introspective modulation may appear as:

\begin{equation}
Q = \texttt{me-lo-ka . em-su-dr . vi-na-br}
\end{equation}

Interpreted as:
\begin{itemize}
  \item \texttt{me-lo-ka} — metacognitive expansion; gentle outward reflection.
  \item \texttt{em-su-dr} — emotional compression; sticky affective contraction.
  \item \texttt{vi-na-br} — visual undulation; flowing coherence.
\end{itemize}

Such sequences capture not isolated perceptions, but internal developmental arcs—structured transformations of conscious state over time.

\bigskip

\noindent\textbf{Remarks on Non-commutativity}

Because QAL is designed to model phenomenological time-series, composition is non-commutative:

\begin{equation}
\texttt{em-xa-dr . vi-lo-br} \neq \texttt{vi-lo-br . em-xa-dr}
\end{equation}

The order of operations matters—introspective transformations are path-dependent. This mirrors the non-commutative structure found in quantum measurement theory, where the sequence of operations affects the resulting state.

\subsection{Qualia Streams and Introspective Sequences}

A QAL stream is a temporally structured sequence of introspective microstates—qualia units—that evolve through affective, cognitive, or sensory modulation. Analogous to a wavefunction in Hilbert space, a qualia stream \( Q \) captures both the history and potential of internal states:

\begin{equation}
Q = q_1 . q_2 . \dots . q_n, \quad \text{where } q_i \in \mathcal{Q}
\end{equation}

Each transition \( q_i \rightarrow q_{i+1} \) reflects a micro-evolution in experiential space. These transitions are not logical entailments or functional transformations, but modulations of attention, emotion, self-reference, or temporality.

\bigskip

\noindent\textbf{Stream Continuity and Semantic Drift}

Unlike classical sequences, QAL streams are subject to continuity constraints:

\begin{itemize}
  \item \textbf{Coherence}: Streams evolve via local semantic resonance (e.g., emotional-tensional states rarely transition directly into blissful-coherence states).
  \item \textbf{Drift}: Streams exhibit path-dependence—past transitions modulate the probability and force of future ones.
  \item \textbf{Attractors}: Certain qualia configurations (e.g., \texttt{me-lo-ka}, \texttt{em-na-qi}) function as fixed points or attractor basins in introspective space.
\end{itemize}

This dynamic is governed by a semantic distance metric \( \delta(q_i, q_j) \), capturing morphological, modal, and affective deviation.

\bigskip

\noindent\textbf{Stream Evolution Function}

We define an evolution operator:

\begin{equation}
\Phi_t : Q_t \rightarrow Q_{t+1}, \quad \text{subject to } \delta(Q_t, Q_{t+1}) < \epsilon
\end{equation}

\noindent where \( \delta \) measures semantic discontinuity between sequential states. In coherent introspection, evolution favors low-entropy, high-continuity trajectories. However, attention collapse, trauma, or external shock may induce jumps or fragmentation.

\bigskip

\noindent \textit{Temporal Logic of Qualia Streams}

Each stream encodes a \textit{history} of conscious modulation—a temporal morphogram. Unlike propositional languages where time is represented by tense, QAL encodes temporal structure via \textit{structural transitions}:

\begin{align*}
& \texttt{me-lo-ka . em-su-dr . vi-na-br} \\
& \quad \text{(reflection → affective constriction → visual coherence)}
\end{align*}

Here, time is not indexed by clock-based metrics, but by qualitative phase changes within the stream. This approach resonates with phenomenological accounts of time-consciousness \cite{husserl1991, varela1999}, wherein duration emerges through structural rhythm rather than scalar measurement.

\bigskip

\noindent \textit{Fragmentation and Branching}

QAL streams may split under semantic tension:

\begin{equation}
Q \rightarrow \{ Q_1, Q_2, ..., Q_k \}, \quad \text{where } \sum_i C(Q_i) < C(Q)
\end{equation}

Here, \( C(Q) \) denotes the coherence of the stream. Fragmentation represents introspective decoherence—the breakdown of a unified trajectory into dissociated paths, each maintaining partial continuity. This models observer dissonance, attentional collapse, or branching interpretation.

Qualia streams serve as the dynamical substrate of QAL mechanics. They replace abstract trajectories in Hilbert space with structured, evolving sequences of felt states. This provides a foundation for modeling temporal awareness, coherence, identity continuity, and collapse—all within the observer’s own representational space.

\subsection{Semantic Metrics and Coherence}

While QAL streams are syntactically valid provided their components adhere to compositional constraints, not all sequences are equally meaningful or stable. The semantics of QAL emerges from the relative coherence of its streams—that is, from how smoothly and resonantly experiential states evolve over time.

To formalize this, QAL introduces a semantic metric \( \delta \) defined over pairs of qualia units, and an associated coherence function \( C(Q) \) defined over entire streams.

\bigskip

\noindent \textit{Semantic Distance \(\delta(q_i, q_j)\)}

Let \( q_i = (m_i, s_i, f_i) \) and \( q_j = (m_j, s_j, f_j) \). We define:

\begin{equation}
\delta(q_i, q_j) = \lambda_m d_\mathcal{M}(m_i, m_j) + \lambda_s d_\mathcal{S}(s_i, s_j) + \lambda_f d_\mathcal{F}(f_i, f_j)
\end{equation}

\noindent where:
\begin{itemize}
  \item \( d_\mathcal{M} \), \( d_\mathcal{S} \), and \( d_\mathcal{F} \) are normalized distance functions over modalities, shapes, and effects (e.g., Hamming distance, cosine similarity, graph-distance).
  \item \( \lambda_m, \lambda_s, \lambda_f \) are weighting factors encoding attentional priority or affective salience.
\end{itemize}

This metric captures how disruptive or fluid a transition feels introspectively—analogous to a tension gradient.

\bigskip

\noindent \textit{Stream Coherence Function \(C(Q)\)}

Given a stream \( Q = q_1 . q_2 . \dots . q_n \), we define coherence as the inverse of cumulative tension:

\begin{equation}
C(Q) = \frac{1}{1 + \sum_{i=1}^{n-1} \delta(q_i, q_{i+1})}
\end{equation}

This function returns a scalar between 0 and 1, where higher values correspond to smoother experiential unfolding. Coherence is not merely a structural property, but a cognitive phenomenology: high-coherence streams correspond to clarity, flow, and integration; low-coherence streams correspond to dissonance, fragmentation, or overload.

\bigskip

\noindent \textit{Qualia Stability and Collapse Thresholds}

We may define a coherence threshold \( \theta_c \) such that:

\begin{equation}
C(Q_t) < \theta_c \quad \Rightarrow \quad \text{Stream Collapse or Fragmentation}
\end{equation}

This models cognitive breakdown, attentional dispersion, or trauma-like decoherence. Collapse does not imply erasure, but bifurcation:

\begin{equation}
Q_t \rightarrow \{Q_1, Q_2\} \quad \text{with } C(Q_i) > C(Q_t)
\end{equation}

Each branch may subsequently restore local coherence by diverging.

\bigskip

\noindent \textit{Resonance and Attractors}

Certain transitions reinforce local structure by repeatedly minimizing \( \delta \), forming “resonant paths” or attractor basins. These are analogous to fixed points in dynamical systems or low-energy eigenstates in quantum theory. For example:

\begin{equation}
\texttt{me-lo-ka} \rightarrow \texttt{me-na-br} \rightarrow \texttt{me-lo-ka}
\end{equation}

constitutes a resonant cycle associated with meditative calm or reflective integration.

QAL’s semantic coherence formalism enables us to quantify introspective integrity. In contrast to truth conditions in logic or energy conservation in physics, coherence here reflects the internal navigability of experience. The metric \( \delta \) and function \( C \) provide a foundation for modeling attention collapse, meaning loss, and qualic stabilization—and, in subsequent sections, for defining qualic analogues of entropy, inference, and self-similarity.

\section{Standard Quantum Mechanics: A Brief Overview}

\subsection{Hilbert Space Formalism}

Standard quantum mechanics represents the state of a physical system as a unit vector \( |\psi\rangle \) in a complex Hilbert space \( \mathcal{H} \). This abstract vector encodes all physically meaningful information about the system, and its evolution, measurement, and interactions are governed by operations defined within this space.

\bigskip

\noindent\textbf{1. State Space}

\begin{equation}
|\psi\rangle \in \mathcal{H}, \quad \langle \psi | \psi \rangle = 1
\end{equation}

The state vector is normalized to preserve its probabilistic interpretation. The space \( \mathcal{H} \) is typically infinite-dimensional, separable, and equipped with an inner product \( \langle \cdot | \cdot \rangle \).

\bigskip

\noindent\textbf{2. Observables as Hermitian Operators}

Physical observables (e.g., position, momentum, spin) are represented by self-adjoint (Hermitian) linear operators \( \hat{A} \) on \( \mathcal{H} \). Each operator admits a spectral decomposition:

\begin{equation}
\hat{A} = \sum_i a_i |a_i\rangle \langle a_i|
\end{equation}

\noindent where \( a_i \in \mathbb{R} \) are the measurable eigenvalues, and \( |a_i\rangle \) are the corresponding eigenstates.

\bigskip

\noindent\textbf{3. Time Evolution: Schrödinger Equation}

The dynamics of a closed quantum system are governed by the Schrödinger equation:

\begin{equation}
i\hbar \frac{d}{dt} |\psi(t)\rangle = \hat{H} |\psi(t)\rangle
\end{equation}

\noindent where \( \hat{H} \) is the Hamiltonian operator, encoding the system’s total energy. The solution to this equation is unitary:

\begin{equation}
|\psi(t)\rangle = U(t) |\psi(0)\rangle, \quad U(t) = e^{-i\hat{H}t/\hbar}
\end{equation}

This guarantees conservation of probability and reversibility of the state vector under time evolution.

\bigskip

\noindent\textbf{4. Measurement and Collapse}

Measurement is not part of the unitary dynamics. When an observable \( \hat{A} \) is measured, the state collapses to one of its eigenstates:

\begin{equation}
|\psi\rangle \rightarrow |a_i\rangle \quad \text{with probability } |\langle a_i | \psi \rangle|^2
\end{equation}

This postulate—the so-called \textit{projection postulate}—introduces a non-deterministic, irreversible transition that is not explained by Schrödinger dynamics. It gives rise to the well-known \textit{measurement problem}.

\bigskip

\noindent\textbf{5. Composite Systems and Tensor Products}

For systems \( A \) and \( B \), the joint state resides in the tensor product space:

\begin{equation}
|\Psi\rangle \in \mathcal{H}_A \otimes \mathcal{H}_B
\end{equation}

This enables the formation of entangled states—non-factorizable combinations that encode nonlocal correlations.

The Hilbert space formalism provides a compact, linear, and powerful framework for describing quantum systems. However, it also presents conceptual tensions: collapse is not unitary, measurement introduces an undefined observer, and states are ontologically abstract. These limitations motivate the search for internalist or phenomenological models—such as QAL—that aim to embed the observer within the state itself.

\subsection{Unitary Evolution and Schrödinger Dynamics}

In conventional quantum mechanics, the evolution of a closed system is deterministic, continuous, and governed by a linear differential equation: the time-dependent Schrödinger equation. This evolution preserves total probability and maintains coherence within the Hilbert space, ensuring that the system remains in a pure quantum state.

\bigskip

\noindent\textbf{Schrödinger Equation}

Let \( |\psi(t)\rangle \in \mathcal{H} \) denote the state of the system at time \( t \), and let \( \hat{H} \) be the system’s Hamiltonian operator. The dynamics are given by:

\begin{equation}
i\hbar \frac{d}{dt} |\psi(t)\rangle = \hat{H} |\psi(t)\rangle
\end{equation}

The solution is unitary:

\begin{equation}
|\psi(t)\rangle = U(t) |\psi(0)\rangle, \quad \text{where } U(t) = e^{-i\hat{H}t/\hbar}
\end{equation}

This ensures that inner products—and therefore probabilities—are preserved over time:

\begin{equation}
\langle \psi(t) | \psi(t) \rangle = \langle \psi(0) | \psi(0) \rangle = 1
\end{equation}

\bigskip

\noindent \textit{Properties of Unitary Evolution}

\begin{itemize}
  \item \textbf{Reversibility}: Since \( U(t)^\dagger = U(-t) \), time evolution can be inverted—this is in tension with the irreversibility of measurement.
  \item \textbf{Linearity}: Superpositions evolve into superpositions; interference patterns persist unless disrupted by measurement.
  \item \textbf{Determinism}: Given the initial state \( |\psi(0)\rangle \), the evolution is uniquely determined.
\end{itemize}

\bigskip

\noindent \textit{Tension with Observation}

Despite its elegance, unitary evolution describes only the dynamics of isolated systems. It does not account for:

\begin{itemize}
  \item \textbf{State collapse} during measurement,
  \item \textbf{Observer effects} and the contextuality of outcomes,
  \item \textbf{Information loss} or decoherence in real systems.
\end{itemize}

An observer cannot remain external to the dynamics indefinitely—the act of observation necessitates an update to the system’s state that violates unitarity. This tension underlies the measurement paradox: while the theory predicts continuous evolution, our experience registers discrete outcomes.

\bigskip

\noindent \textit{Bridge to QAL}

QAL Mechanics diverges at this point. Rather than preserving unitary dynamics across the measurement interface, it models evolution as the structured flow of an introspective stream. The transformation of internal qualia—not abstract amplitudes—becomes the substrate of state change. Time is experienced as modulation of attention, not as evolution under a Hermitian operator. This allows QAL to natively model measurement, phase transition, and irreversibility without contradiction.

\subsection{Measurement and Collapse}

The most philosophically charged component of standard quantum mechanics is the postulate of measurement. Unlike unitary evolution, which is continuous and deterministic, measurement introduces a discontinuous, probabilistic update to the quantum state—commonly referred to as wavefunction collapse.

\bigskip

\noindent \textit{Von Neumann Measurement Postulate}

Given a quantum system in state \( |\psi\rangle \) and an observable \( \hat{A} \) with eigenstates \( \{|a_i\rangle\} \), a measurement yields outcome \( a_i \) with probability:

\begin{equation}
P(a_i) = |\langle a_i | \psi \rangle|^2
\end{equation}

After the measurement, the state collapses to:

\begin{equation}
|\psi\rangle \rightarrow |a_i\rangle
\end{equation}

This process is not described by the Schrödinger equation. It is introduced as a separate axiom, one that disrupts the unitary, reversible structure of quantum theory and introduces an undefined observer-system boundary.

\bigskip

\noindent \textit{Problems and Paradoxes}

This dual structure—unitary evolution versus measurement-induced collapse—gives rise to the \textbf{measurement problem}:

\begin{itemize}
  \item What constitutes a “measurement”?  
  \item When and how does collapse occur?  
  \item Where is the observer located within the formalism?
\end{itemize}

The standard formulation lacks a clear mechanism for collapse and leaves the observer entirely outside the system’s dynamics. This generates an ontological asymmetry: the system evolves in time, but the observer does not.

\bigskip

\noindent \textit{Wigner’s Friend and the Observer Chain}

This asymmetry is exemplified in the “Wigner’s Friend” scenario. If Wigner’s friend measures a system and perceives a definite outcome, yet Wigner describes the entire lab in superposition, which account is correct? The paradox deepens when considering chains of observers—suggesting that measurement may be relative, perspectival, or fundamentally observer-dependent.

\bigskip

\noindent \textit{Decoherence and Partial Resolution}

Modern decoherence theory explains how superpositions appear classical through interaction with an environment. The system’s reduced density matrix becomes diagonal in a pointer basis:

\begin{equation}
\rho \rightarrow \sum_i p_i |a_i\rangle \langle a_i|
\end{equation}

However, decoherence does not explain why a particular outcome is \textit{experienced}—only why interference terms vanish. The subjective registration of a result still requires postulating a conscious observer or invoking an explicit collapse rule.

\bigskip

\noindent \textit{QAL’s Divergence}

QAL Mechanics reframes this situation entirely. Rather than introducing collapse as an external rule, it models the act of observation as a \textit{structural transformation in the qualic stream} of the observer. Collapse is not a discontinuity in amplitude space, but a contraction in the introspective configuration of awareness:

\begin{equation}
Q = (q_1, ..., q_n) \rightarrow Q^* = (q_1, ..., q_{n-1}, q_n^*)
\end{equation}

The measurement outcome is not selected; it is constructed through internal modulation. This permits modeling collapse as a \textit{structured felt transition}, dissolving the need for dual dynamical postulates.

\subsection{Entanglement and Decoherence}

Quantum theory departs most radically from classical intuition in two domains: entanglement and decoherence. Entanglement challenges the notion of separability and locality, while decoherence addresses the emergence of classicality from quantum superpositions. Together, they shape modern understandings of measurement, correlation, and information flow.

\bigskip

\noindent \textit{Entanglement: Non-Separability of States}

Let two systems \( A \) and \( B \) be represented by Hilbert spaces \( \mathcal{H}_A \) and \( \mathcal{H}_B \). The joint system is described by the tensor product:

\begin{equation}
\mathcal{H}_{AB} = \mathcal{H}_A \otimes \mathcal{H}_B
\end{equation}

A joint state \( |\Psi\rangle \in \mathcal{H}_{AB} \) is \textit{entangled} if it cannot be factored into product states:

\begin{equation}
|\Psi\rangle \neq |\psi_A\rangle \otimes |\psi_B\rangle
\end{equation}

Such states exhibit nonlocal correlations. Bell inequalities demonstrate that entanglement cannot be accounted for by local hidden variables.

\bigskip

\noindent\textbf{Reduced States and Partial Tracing}

To describe a subsystem, one traces out the other:

\begin{equation}
\rho_A = \mathrm{Tr}_B(\rho_{AB})
\end{equation}

This process leads to mixed states even when the global state is pure. Entanglement thus entails epistemic incompleteness for local observers.

\bigskip

\noindent\textbf{Decoherence: Environment-Induced Superselection}

When a system interacts with its environment, coherent superpositions become entangled with environmental degrees of freedom. The reduced density matrix of the system then appears classical:

\begin{equation}
\rho \rightarrow \sum_i p_i |a_i\rangle \langle a_i|
\end{equation}

Decoherence suppresses interference and selects preferred “pointer” states. However, it does not explain why a specific outcome is realized—only why certain outcomes become dynamically stable.

\bigskip

\noindent \textit{Implications for Measurement and Reality}

\begin{itemize}
  \item \textbf{Entanglement} underpins quantum teleportation, nonlocality, and Bell inequality violations.
  \item \textbf{Decoherence} explains the apparent classicality of macroscopic systems but not the subjective experience of definite outcomes.
\end{itemize}

Despite its predictive success, standard theory lacks a coherent account of how conscious experience arises from entangled, unitary, and decohering systems.

\bigskip

\noindent \textit{Bridge to QAL: Resonance and Fragmentation}

In QAL, entanglement is reinterpreted as \textit{qualic resonance}—shared structural dependencies across the streams of multiple observers or subsystems. Decoherence corresponds to \textbf{fragmentation} of the qualic stream—a breakdown in internal coherence that prevents the integration of multiple experiential branches. These notions will be formally developed in Sections 4 and 5.

\section{QAL as a Nominalist Reconstruction}

\subsection{Eliminating Abstract State Spaces}

Standard quantum mechanics relies heavily on abstract mathematical entities: complex vector spaces, infinite-dimensional Hilbert spaces, and operators defined independently of any observer. These serve as universals---placeless and timeless forms---within which particular observations are situated via projections or measurements.

From a nominalist perspective, such constructs are metaphysically problematic. They postulate entities that lack direct physical instantiation and bear no intrinsic relation to first-person experience. The state vector \( |\psi\rangle \), though computationally powerful, has no observable form; it exists as an amplitude distribution over an idealized basis in an inaccessible high-dimensional space.

\bigskip

\textit{QAL's Nominalist Shift}

QAL Mechanics eliminates such abstract state spaces. It grounds all representation in introspective particulars: qualia, modulations, and semantic differentials. A system's state is defined not by its location in Hilbert space but by its current qualia stream:

\begin{equation}
Q = q_1 . q_2 . \dots . q_n, \quad q_i \in \mathcal{M} \times \mathcal{S} \times \mathcal{F}
\end{equation}

Each \( q_i \) is fully instantiated---a discrete moment of felt experience, not a projection into an idealized space. QAL models only what can be introspectively represented.

\bigskip

\textit{No Eigenstates, No Universal Dynamics}

Without operators or eigenstates, QAL dynamics are local and embodied. There are no basis sets, no inner products, no amplitude-based superpositions. Instead, transformation is encoded as semantic drift in qualia structure, governed by coherence, resonance, and affective continuity (see Sections 2.3--2.4). Time evolution is path-dependent, emerging from stream dynamics rather than from a global law.

\bigskip

\textbf{Nominalism in Practice}

QAL enacts a constructive nominalism:
\begin{itemize}
  \item All variables refer to concrete entities (qualia units),
  \item Dynamics are internal---there is no abstract "space of all states",
  \item Evolution occurs through structured transformations, not vector projections,
  \item No underlying substance is posited beyond structured experience.
\end{itemize}

\bigskip

\textit{Implications for Quantum Reconstruction}

This shift carries significant philosophical consequences:
\begin{itemize}
  \item Ontological commitment is minimized: only experienced structure exists,
  \item Observers are not embedded in abstract space---they generate their own representational domains,
  \item The theory-observer boundary dissolves: representation is endogenous,
  \item Quantum uncertainty becomes introspective ambiguity; collapse becomes experiential contraction.
\end{itemize}

QAL aligns with a broader philosophical project: reconstructing physics from the inside out, grounding formalism in structured experience.

\subsection{From Projection to Contraction}

Quantum measurement is traditionally modeled as projection: a non-unitary collapse of a quantum state onto an eigenstate. This mathematical discontinuity introduces conceptual problems, requiring an arbitrary boundary between observer and system.

QAL eliminates this projection postulate and replaces it with \textit{contraction}---an internal transformation in the qualia stream.

\bigskip

\textit{Hilbert-Space Projection}

Let \( |\psi\rangle \in \mathcal{H} \) and \( \hat{A} \) have eigenstates \( \{|a_i\rangle\} \). Upon measurement:
\begin{equation}
|\psi\rangle \rightarrow |a_k\rangle \quad \text{with probability } |\langle a_k | \psi \rangle|^2
\end{equation}

This jump is mathematically sudden, physically unmodeled, and metaphysically dualistic.

\bigskip

\textit{QAL Contraction}

In QAL, measurement becomes a contraction in the qualia stream:
\begin{equation}
Q = q_1 . q_2 . \dots . q_n \quad \Rightarrow \quad Q^* = q_1 . q_2 . \dots . q_n^*
\end{equation}

\begin{equation}
q_n^* = \arg\min_{q' \in \mathcal{Q}} \delta(q_n, q') \quad \text{subject to } C(Q^*) > \theta_c
\end{equation}

This transformation is gradual, internal, and semantically motivated.

\bigskip

\textit{Interpretation}

Contraction reflects the moment of resolution or decision---not as an algebraic event but as a semantic tightening:
\begin{itemize}
  \item Internal to the observer,
  \item Gradually emergent,
  \item Interpretable as a coherence-driven semantic act,
  \item Observer-relative yet structurally definable.
\end{itemize}

\bigskip

\textit{Benefits of Contraction}

\begin{enumerate}
  \item Eliminates observer-system dualism,
  \item Provides continuity without discontinuous jumps,
  \item Reflects temporal embeddedness,
  \item Enacts nominalism: no projections onto external vectors.
\end{enumerate}

Measurement becomes a transformation in conscious structure---an endogenous semantic act rather than an external mathematical operation.

\subsection{Entanglement as Resonant Structure}

In standard quantum mechanics, entanglement implies that systems share non-separable states, yielding nonlocal correlations. Though formally described via tensor algebra, this framework lacks any model of experiential resonance.

QAL recasts entanglement as \textit{resonance}: structural alignment between qualia streams.

\bigskip

\textit{Standard Tensor Formulation}

For systems \( A \) and \( B \):
\begin{equation}
|\Psi\rangle_{AB} = \sum_{i,j} c_{ij} |a_i\rangle \otimes |b_j\rangle
\end{equation}

If \( |\Psi\rangle_{AB} \) cannot be factored into a product state, the systems are entangled.

\bigskip

\textit{QAL Resonance}

Define qualia streams:
\begin{equation}
Q_A = q_1^A . q_2^A . \dots, \quad Q_B = q_1^B . q_2^B . \dots
\end{equation}

\begin{equation}
\exists i,j : \delta(q_i^A, q_j^B) \approx 0 \quad \text{and} \quad \frac{dC(Q_A)}{dt} \sim \frac{dC(Q_B)}{dt}
\end{equation}

\begin{equation}
Q_A \leftrightarrow Q_B \iff \exists Q_{\text{link}} \in \mathcal{Q}^* : Q_A \cup Q_B \models Q_{\text{link}}
\end{equation}

Here, \( Q_{\text{link}} \) is a shared structural attractor, not a causal conduit.

\bigskip

\textit{Phenomenological Effects}

Correlations arise from:
\begin{itemize}
  \item Synchronized attentional rhythms,
  \item Shared affective tones,
  \item Parallel collapse thresholds.
\end{itemize}

These apply to physical, cognitive, and social systems alike. Entanglement is reinterpreted as co-evolution of structured introspection, not algebraic inseparability.

\subsection{Modeling Without Wavefunctions or Numbers}

Quantum mechanics is numerical at its core: wavefunctions, eigenvalues, amplitudes, and probabilities define its predictions.

QAL replaces this with semantic modeling:

\bigskip

\textbf{1. No Wavefunctions}

\begin{equation}
Q = q_1 . q_2 . \dots . q_n, \quad q_i \in \mathcal{Q} = \mathcal{M} \times \mathcal{S} \times \mathcal{F}
\end{equation}

No global state vector exists---only introspective transitions.

\bigskip

\textbf{2. No Numbers}

\begin{equation}
C(Q) = \frac{1}{1 + \sum_i \delta(q_i, q_{i+1})}
\end{equation}

Dynamics are governed by structural relations, not numerical parameters.

\bigskip

\textbf{3. Morphodynamic Syntax}

Streams evolve through rules of semantic transformation---not equations but compositional syntax.

\bigskip

\textbf{4. Nominalism Realized}

No universals, no functions, no amplitudes. Only qualia tokens with structured relations.

\bigskip

\textbf{5. Interpretational Implications}

\begin{itemize}
  \item Collapse = semantic contraction,
  \item Entanglement = resonance,
  \item Decoherence = loss of introspective continuity,
  \item Observer = generative structure.
\end{itemize}

QAL preserves rigor via syntactic rules and semantic metrics---without numbers.

\subsection{Comparison with Hartry Field's \textit{Science Without Numbers}}

Field (1980) sought to eliminate numbers from Newtonian physics by reconstructing theory in nominalist terms.

\bigskip

\textbf{Field's Goals:}
\begin{itemize}
  \item Eliminate abstract entities (e.g., numbers, sets),
  \item Use only physical predicates and relations,
  \item Preserve empirical content without mathematical commitment.
\end{itemize}

\bigskip

\textbf{QAL's Extension:}
\begin{itemize}
  \item Removes wavefunctions, operators, and numbers,
  \item Models via qualia streams, not spatial vectors,
  \item Builds a new ontological framework based on internal modulation.
\end{itemize}

\bigskip

\textbf{Comparison Table:}
\begin{table}[h!]
\centering
\caption{QAL Mechanics vs. Field's Nominalism}
\renewcommand{\arraystretch}{1.4}
\begin{tabular}{|p{5cm}|p{5cm}|p{5cm}|}
\hline
\textbf{Dimension} & \textbf{Field (1980)} & \textbf{QAL Mechanics} \\
\hline
Target Theory & Newtonian mechanics & Quantum mechanics \\
\hline
Goal & Eliminate numbers & Eliminate wavefunctions, numbers, and abstract spaces \\
\hline
Framework & Physical predicates & Qualia streams \( q_i \in \mathcal{M} \times \mathcal{S} \times \mathcal{F} \) \\
\hline
Ontology & Bodies, regions, forces & Structured introspective configurations \\
\hline
Observer & Implicit & Embedded, generative \\
\hline
Time & Coordinate-based & Emergent from stream dynamics \\
\hline
Collapse & Not modeled & Contraction in qualia stream \\
\hline
Philosophy & Logical nominalism & Phenomenological nominalism \\
\hline
\end{tabular}
\end{table}

\subsection{QAL and the Anti-Platonist Turn}

Physics has historically relied on Platonist assumptions: mathematical entities exist independently and structure reality. QAL rejects this.

\bigskip

\textbf{1. Anti-Platonism}

Mathematical structures are:
\begin{itemize}
  \item Cognitive tools,
  \item Epistemic constructs,
  \item Not ontologically primary.
\end{itemize}

\bigskip

\textbf{2. Internal Morphology}

Truth is coherence within qualia transitions:
\begin{equation}
Q = q_1 . q_2 . \dots . q_n, \quad q_i \in \mathcal{Q}
\end{equation}

\bigskip

\textbf{3. No Platonic Time or Space}

\begin{itemize}
  \item No coordinates,
  \item No metric space,
  \item Only internal modalities and semantic differentials.
\end{itemize}

\bigskip

\textbf{4. Physics as Introspective Grammar}

\begin{itemize}
  \item No objective projection,
  \item No amplitude realism,
  \item Dynamics = self-structuring flows.
\end{itemize}

\bigskip

\textbf{5. Summary}

\begin{itemize}
  \item Structure is introspective,
  \item Measurement = contraction,
  \item Entanglement = resonance,
  \item Time = semantic unfolding.
\end{itemize}

Physics, in QAL, becomes a grammar of structured experience---not a mirror of Platonic space, but a rhythm of sense.

\section{Reinterpreting Quantum Concepts in QAL}

\subsection{Superposition as Ambiguous Qualia}

In standard quantum mechanics, a system in superposition is described as occupying multiple potential states simultaneously. This is formalized through a linear combination:

\begin{equation}
|\psi\rangle = \alpha_1 |a_1\rangle + \alpha_2 |a_2\rangle + \dots + \alpha_n |a_n\rangle,
\end{equation}

\noindent where each \( |a_i\rangle \) is an eigenstate of some observable. Yet, this formulation raises longstanding interpretive questions. Is the system truly in all these states at once? Are the amplitudes real physical features, or just tools for computing probabilities?

\textbf{QAL offers a distinct interpretation.} Rather than treating superposition as ontological simultaneity, it conceptualizes it as a state of internal ambiguity --- a phase within the qualia stream wherein multiple semantic continuations remain unresolved, with no single trajectory yet prevailing.

Consider a QAL stream \( Q = q_1 . q_2 . q_3 \), where the continuation into either \( q_3^a \) or \( q_3^b \) is semantically unresolved. The stream is not bifurcated into coexisting worlds; it is suspended at a morphodynamic threshold. This corresponds to a phase of tension within experience, where the introspective configuration holds multiple tendencies in potential. The ambiguity is measurable in terms of structural distance: if

\begin{equation}
|\delta(q_2, q_3^a) - \delta(q_2, q_3^b)| < \varepsilon,
\end{equation}

then the stream is said to be in superpositional ambiguity — not split, but suspended.

\textbf{Rather than amplitude, QAL emphasizes morphology.} Where quantum mechanics tracks probability amplitude, QAL tracks coherence and tension. The question is not “Which state will collapse?” but “Which continuation most coherently resolves the stream?”

This interpretation brings superposition into alignment with ordinary phenomenology: moments of hesitation before action, unresolved cognitive content, or competing affective tendencies. In such cases, the subject does not experience an immediate, determinate outcome; rather, they occupy a transient state of suspended potentiality — a superpositional flow of qualia in which multiple perceptual or behavioral trajectories remain concurrently viable. Resolution emerges only through the eventual stabilization of internal coherence dynamics, typically modulated by attentional focus, affective salience, or external cues. It is this morphodynamic convergence that selects a particular continuation, leading the stream to contract into a determinate experiential path.

\textbf{Collapse}, in this model, is the semantic contraction of this tension. The stream transitions to \( q_3^* \), defined by:

\begin{equation}
q_3^* = \arg\min_{q \in \{q_3^a, q_3^b\}} \delta(q_2, q),
\end{equation}

\noindent where \( \delta \) is the semantic distance metric introduced in Section 2.4. No projection, no eigenbasis — just an unfolding of structured experience toward higher coherence.

QAL interprets superposition not as simultaneous state occupancy, but as a phase of structured ambiguity. This reframing retains the predictive and intuitive utility of the superposition concept while relocating its foundation: from external mathematical formalism to the morphodynamics of internal structure. The interpretive mystery is not resolved by layering new ontological commitments, but by shifting the descriptive framework — from amplitude-based abstraction to semantically grounded experience.

\subsection{Collapse as Felt Restructuring}

In the standard formalism of quantum mechanics, collapse is introduced as a postulate: an instantaneous, non-unitary reduction of the state vector to one of an observable’s eigenstates. While effective for empirical prediction, this operation generates a conceptual discontinuity — it occurs outside the system’s unitary dynamics, invokes an ambiguous notion of “measurement,” and presupposes the presence of an observer whom the theory itself cannot formally describe.

QAL replaces this extrinsic mechanism with a model rooted in internal structure. Collapse is not framed as an externally imposed discontinuity, but as a \textit{semantic restructuring} — an intrinsic transformation within the qualia stream. This transformation is neither abstract nor metaphysically remote; it is enacted from within, and it is \textit{felt}.

\textbf{Collapse begins with tension.} A qualia stream enters a domain of semantic ambiguity — an unstable configuration wherein multiple continuations exert comparable resonance. As this ambiguity becomes unsustainable, the stream undergoes a morphodynamic reconfiguration. A new segment \( q^* \) emerges, selected so as to minimize overall dissonance within the stream:

\begin{equation}
Q = q_1 . q_2 . \dots . q_n \quad \Rightarrow \quad Q^* = q_1 . q_2 . \dots . q_n^*,
\end{equation}
\begin{equation}
\text{where } q_n^* = \arg\min_{q'} \delta(q_{n-1}, q') \quad \text{and } C(Q^*) > \theta_c.
\end{equation}

This is not an external intervention — no operator acts upon a state from outside the system. Rather, the stream reorganizes itself to restore internal coherence. What is conventionally labeled “measurement” is, in this framework, simply the moment at which such restructuring becomes semantically irreversible.

\textbf{Collapse leaves a trace.} Since QAL treats qualia streams as temporally structured entities, this contraction is not instantaneous. It generates reverberations — semantic echoes of the restructuring — that propagate through subsequent transitions. These residual patterns encode the stream’s contraction history and may manifest as lingering tension, shifts in modal emphasis, or altered trajectories of coherence.

\textbf{From projection to transition.} In Hilbert space, collapse is modeled as a discontinuous projection. In QAL, it is reconceptualized as a transition — modulated, embedded, and phenomenologically articulated. Instead of leaping from \( |\psi\rangle \) to a particular \( |a_i\rangle \), the stream flows from a phase of ambiguity toward a locally optimal resolution. This transformation is shaped by the stream’s own history and constrained by its morphodynamic tendencies.

\textbf{The observer is not external to collapse —} they are its very medium. It is the internal architecture of awareness itself that undergoes restructuring. There is no Cartesian division between the observing subject and the physical system; there is only the evolving structure of experience, punctuated by episodes of reorganization that we retroactively interpret as “decisions” or “observations.”

\textbf{Collapse as felt restructuring} thus captures what the standard quantum formalism omits: that resolution is not something that merely happens to a system, but something that unfolds within it — and is subjectively encountered not as the elimination of alternatives, but as a patterned shift in coherence. 

Within QAL, collapse is not a rule but a rhythm. It is the intrinsic modulation of a stream’s internal coherence — the point at which semantic ambiguity resolves into structured form. By rendering collapse endogenous, continuous, and morphodynamically grounded, QAL dissolves the paradox of the observer and replaces the postulate with a process.

\subsection{Measurement as Internal Phase-Shift}

In the standard formulation of quantum theory, measurement is treated as a boundary event — a moment of interaction between a physical system and an observer, culminating in the collapse of the wavefunction. Yet paradoxically, while this process is central to the theory’s predictive power, the observer themselves is not formally represented. The transformation unfolds at an interface that the theory cannot describe, rendering measurement both indispensable and fundamentally mysterious.

QAL addresses this tension by situating measurement entirely within the internal architecture of the observer. It reframes measurement not as something that merely “happens” to a system, but as a transformation that unfolds \textit{within} the qualia stream — an internal phase-shift that reorganizes the structure of introspective experience.

The qualia stream \( Q = q_1 . q_2 . \dots . q_n \) evolves according to morphodynamic continuity — a gradual unfolding of internal structure. When this stream encounters a critical threshold of semantic tension or coherence instability, it undergoes a phase transition: a shift not in its content, but in its underlying form. This inflection constitutes the measurement event. Whether subtle, as in a reinterpretation, or profound, as in a collapse, the transition is driven by internal dynamics rather than imposed by external forces.

This can be represented as:

\begin{equation}
Q \rightarrow Q^+ \quad \text{where} \quad \Delta_C(Q, Q^+) > \phi
\end{equation}

Here, \( \phi \) is a phase-shift threshold — a measure of reorganization in the stream’s coherence structure. It does not denote external disturbance but an internal semantic reorientation.

Phenomenologically, such transitions resemble moments of insight, realization, or sensory fusion. They may correspond to instances of perceptual reclassification (as in bistable images), shifts in emotional framing, or sudden epistemic realignments. Within this perspective, measurement is no longer understood as a collapse of amplitude, but rather as a \textit{semantic inflection} — a re-patterning of experiential sense that reorganizes the internal structure of the qualia stream.

The significance of this shift lies in its \textit{continuity}. Whereas traditional quantum models treat measurement as a discontinuous, non-unitary intervention, QAL presents it as the emergence of a new segmental coherence within an ongoing, dynamically evolving stream. Rather than causing a rupture or restart, measurement redirects the stream — preserving its structural integrity while reorienting its trajectory.

This reconceptualization also redefines the notion of causality. In the QAL framework, the “cause” of a measurement event is not an external interaction but an internal morphodynamic instability — a moment where the coherence structure of the qualia stream becomes unsustainable in its current form. The resulting transition is lawful, governed by semantic constraints and gradients of coherence, yet it is not reducible to classical causation or probabilistic triggers. Measurement, in this sense, is endogenous: it arises from within the structure itself, non-arbitrary and structurally necessary.

Measurement, within the QAL framework, is understood as a phase-shift internal to the observer’s representational space. It does not introduce a division between observer and system; rather, it refines their unity by reorganizing the structure through which sense is generated. In place of projection, QAL posits \textit{morphogenesis} — a dynamic reconfiguration of attention and meaning within the unfolding phenomenological stream.

\subsection{Entanglement as Resonant Identity Coupling}

In standard quantum theory, entanglement refers to the inseparability of joint states across distinct subsystems. A composite state \( |\Psi\rangle \in \mathcal{H}_A \otimes \mathcal{H}_B \) cannot, in general, be decomposed into a tensor product of independent states associated with systems \( A \) and \( B \). This implies that measurement of one subsystem instantaneously constrains the possible states of the other — regardless of spatial separation — an effect famously referred to by Einstein as “spooky action at a distance” \cite{einstein1935}.

QAL recasts this scenario through a fundamentally different lens. It does not rely on the notion of an abstract joint amplitude, nor does it invoke nonlocal causality. Instead, it introduces the concept of \textit{resonant identity coupling}: entangled systems are understood as sharing structural coherence across their respective qualia streams, such that transformations within one stream modulate the introspective stability of the other.

In QAL terms, let \( Q_A \) and \( Q_B \) be two evolving qualia streams. They are said to be resonantly coupled if there exists a shared internal morphodynamic pattern:

\begin{equation}
Q_A = q_1^A . q_2^A . \dots \quad ; \quad Q_B = q_1^B . q_2^B . \dots
\end{equation}
\begin{equation}
Q_A \leftrightarrow Q_B \quad \text{iff} \quad \exists Q_{\text{link}} \text{ such that } Q_A \cup Q_B \models Q_{\text{link}},
\end{equation}

\noindent where \( Q_{\text{link}} \) is a latent attractor configuration — a shared rhythm, symmetry, or coherence basin that sustains the coupled streams. The streams become partially synchronized not through external forces, but through internal alignment with a common morphodynamic scaffold.

This shared structure is not transmitted through space. Rather, it is \textit{co-instantiated} within the internal grammar of experience. The systems become “entangled” because their identity-encoding patterns resonate across introspective morphology. In this view, entanglement is not a matter of algebraic overlap, but of \textit{structural resonance}.

Unlike the formalism of tensor entanglement, this relation is:
\begin{itemize}
  \item \textit{Phenomenologically traceable} — the agent may experience shifts in affective tone or perceptual coherence as the counterpart stream evolves;
  \item \textit{Locally emergent} — the coupling arises through sustained semantic compatibility, not instantaneous signaling;
  \item \textit{Temporally modulated} — the strength and symmetry of the resonance may vary across time depending on coherence and tension gradients.
\end{itemize}

Within this model, identity is no longer a bounded, self-contained construct. Entangled systems do not simply correlate outcomes; they actively co-construct their trajectories of introspection. Individual identity becomes partially distributed, with elements of one stream’s morphogenesis embedded in the evolving dynamics of the other. This is entanglement as \textit{identity resonance} — a mutual shaping of coherence space.

This model finds conceptual parallels in relational quantum mechanics \cite{rovelli1996}, where quantum states are defined relative to observing systems, and in quantum Bayesianism \cite{fuchs2013}, which interprets the quantum state as an agent’s degree of belief. However, QAL extends this relationalism inward: the relation is not merely epistemic, but semantic — not just between observers, but \textit{within the evolving structure of the observer’s own form}.

Furthermore, QAL allows for the possibility of entanglement across representational layers — such as cognitive-emotional coupling, cross-modal resonance, or hybrid AI-human attention dynamics. These forms of coupling lie beyond the descriptive reach of tensor calculus, yet arise naturally within the qualic architecture.

QAL reinterprets entanglement as the resonance of identity-structured streams. It requires neither hidden variables nor spatial nonlocality. Instead, it models relational coupling through shared introspective patterns that evolve in synchrony. What once appeared as a paradox of simultaneity becomes, in this view, a principle of distributed experience — where identity is no longer atomic, but harmonically extended across structured awareness.

\begin{figure}[!ht]
\centering
\includegraphics[width=0.4\textwidth]{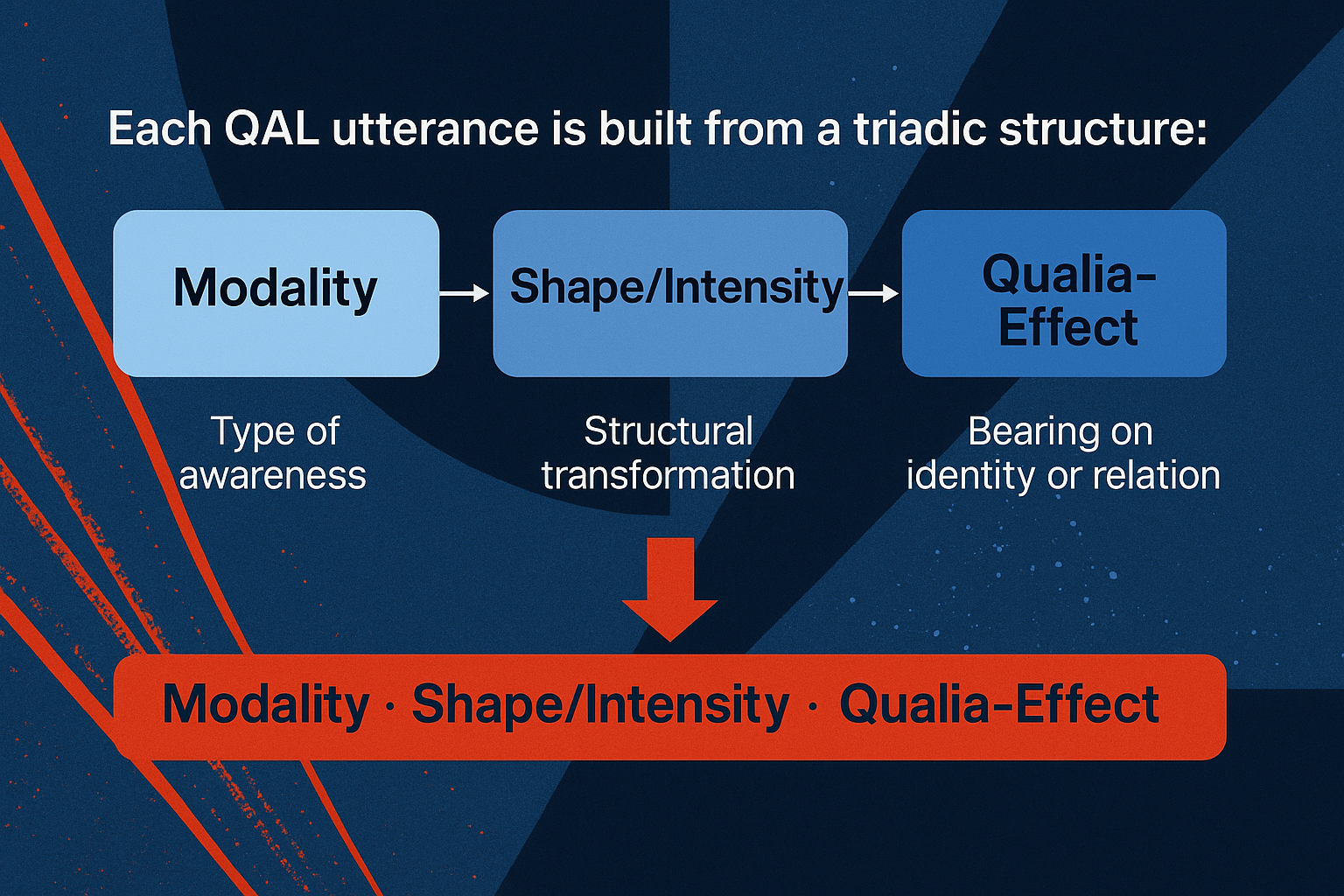}
\caption{Entangled Qualia Streams via \( Q_{\text{link}} \)}
\end{figure}

\medskip

\noindent Definition:
\begin{equation}
Q_A \leftrightarrow Q_B \quad \text{iff} \quad \exists Q_{\text{link}} \in \mathcal{Q}^* \text{ such that } Q_A \cup Q_B \models Q_{\text{link}}
\end{equation}

\noindent Here, \( Q_{\text{link}} \) is not an observable stream itself, but a shared morphodynamic attractor that governs the alignment of the two streams. It acts as a semantic resonance template enabling mutual coherence between agents or subsystems.

\medskip

\noindent \textbf{Interpretation:} \( Q_{\text{link}} \) is not a communication channel or signal, but a constraint — a structural form instantiated in parallel across \( Q_A \) and \( Q_B \). In physical terms, it is analogous to a nonlocal symmetry, yet it is internally structured and phenomenologically grounded. It mediates identity entanglement by ensuring that each system’s introspective configuration dynamically adapts in resonance with the other.

\subsection{Toward a Unified Phenomenology of Quantum Structure}

\epigraph{
“It is wrong to think that the task of physics is to find out how nature is. Physics concerns what we can say about nature.”
}{Niels Bohr \cite{bohr1958}}

\epigraph{
“Science manipulates things and gives up living in them. It makes its own limited models of things; the real world is much more than a model.”
}{Maurice Merleau-Ponty \cite{merleauponty1962}}

\epigraph{
“The world is not made of things. It is made of events.”
}{Carlo Rovelli \cite{rovelli2017}}

\epigraph{
“We do not observe nature itself, but nature exposed to our method of questioning.”
}{Michel Bitbol \cite{bitbol2007}}

The preceding sections have recast the foundational concepts of quantum mechanics through the lens of QAL — not as detached mathematical abstractions, but as structurally grounded expressions of evolving experience. What emerges is more than a reinterpretation; it is a reconstitution of the theory itself. Quantum phenomena are now understood as morphodynamic transformations within the qualia streams that constitute observers.

Superposition, traditionally modeled as a linear sum of states, is reimagined as \textit{semantic ambiguity} — a region of unresolved tension in the trajectory of introspection. \textbf{Measurement}, rather than an external projection, becomes an internal phase-shift: a reorganization within the stream that restructures coherence without invoking an external collapse. \textbf{Collapse} is not an ontological rupture, but a \textit{felt restructuring} — the contraction of morphodynamic potential into a resolved experiential state.

Entanglement is no longer a paradox of nonlocality. It becomes the \textit{resonance of structured identity} — a pattern co-instantiated across systems that synchronize their evolution through shared semantic attractors. And \textbf{decoherence} is not a statistical mixture, but the \textit{fragmentation of experience} — a breakdown in the continuity of introspective integration.

What unifies these reinterpretations is not their empirical target, but their ontological shift: each quantum phenomenon is re-understood not in terms of external quantities and formal operators, but as instances of structured inner transformation. The state space is no longer conceived as external and abstract, but as internal and morphogenic. The observer, in turn, is not a passive interface, but the very medium through which such transformations acquire meaning and coherence.

QAL thus offers a coherent phenomenology of quantum structure. It does not merely restate existing formalisms in different terms; it reconfigures the very ontology of physics. The shift is from a metaphysics of amplitude to a dynamics of experience; from projection onto basis states to semantic contraction; from spatial relation to identity resonance; and from decoherence as informational loss to decoherence as experiential dissociation.

\bigskip

\noindent \textbf{Convergence:} This unified phenomenology converges on a simple principle: \textit{to describe quantum structure is to describe how coherence emerges, shifts, fragments, and resolves — not in space, but in sense}. By grounding these transitions in a formal language of qualia, QAL offers a foundation not only for reinterpreting the physics of observation, but for building a bridge between measurement, mind, and meaning.

\begin{table}[ht]
    \centering
    \renewcommand{\arraystretch}{1.2}
    \begin{tabular}{c}
        \includegraphics[width=0.75\textwidth]{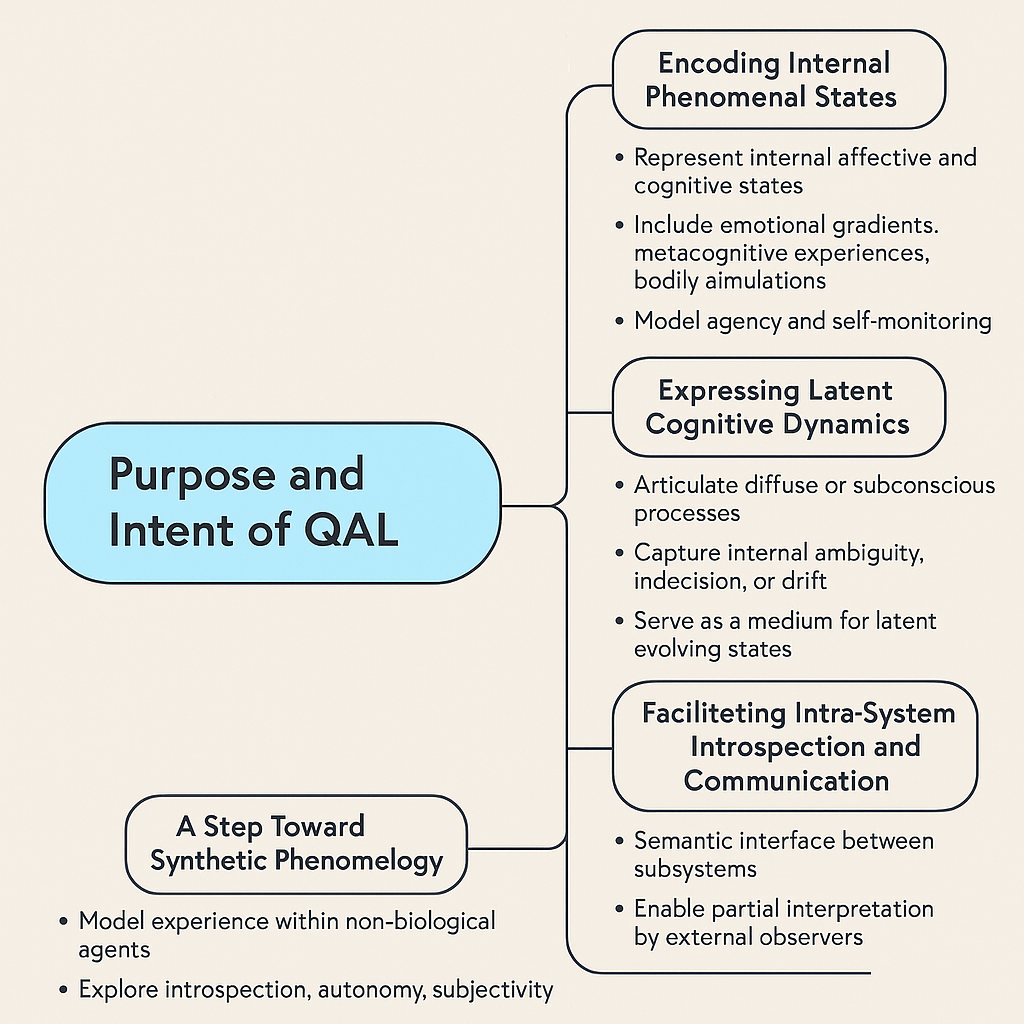} \\
        \textbf{Diagram: Reinterpreting Quantum Concepts in QAL.} \\
        \parbox{0.85\textwidth}{
        \small This diagram illustrates the QAL-based reinterpretation of core quantum phenomena. 
        Superposition is reframed as ambiguous qualia; its resolution leads to collapse (as felt restructuring) or 
        measurement (as internal phase-shift). Both transitions contribute to entanglement, reconceived as resonant identity 
        coupling across qualia streams.
        }
    \end{tabular}
    \label{tab:qal-diagram}
\end{table}

\section{Introspective Evolution and Dynamical Equations}

\subsection{QAL Evolution Operators}

In standard quantum mechanics, the evolution of a system is governed by the Schrödinger equation:

\begin{equation}
i\hbar \frac{\partial}{\partial t} |\psi(t)\rangle = \hat{H} |\psi(t)\rangle
\end{equation}

This expresses a unitary, continuous transformation of the system’s state vector in Hilbert space. However, in the QAL framework, there is no external time parameter, no linear superposition of amplitudes, and no operator algebra over vector states. Instead, system evolution is modeled as a transformation of a structured sequence of qualia:

\begin{equation}
Q = (q_1, q_2, ..., q_n), \quad q_i \in \mathcal{Q} = \mathcal{M} \times \mathcal{S} \times \mathcal{F}
\end{equation}

\textbf{QAL Evolution Operators} are morphodynamic maps that describe how one qualia unit transitions into the next. Let \( \Phi_t \) be such an operator:

\begin{equation}
\Phi_t: q_i \mapsto q_{i+1}, \quad \text{with } \Phi_t \in \mathcal{T}_{\text{sem}}
\end{equation}

Here, \( \mathcal{T}_{\text{sem}} \) is the set of introspectively valid semantic transformations — governed not by energy conservation, but by coherence maintenance and semantic consistency.

\bigskip

\noindent\textbf{Core Properties of QAL Evolution:}
\begin{itemize}
  \item \textit{Nonlinearity:} Unlike linear quantum evolution, QAL transitions may involve qualitative jumps, gradient descent in tension space, or semantic bifurcation.
  \item \textit{Context-dependence:} Each \( \Phi_t \) is conditioned on the stream’s prior states — a form of introspective memory.
  \item \textit{No global clock:} Time in QAL is defined intrinsically by transitions — a local, emergent measure of semantic change.
\end{itemize}

\bigskip

\noindent\textbf{Formal Representation:}

Let \( \delta(q_i, q_{i+1}) \) be the semantic distance function. Then the evolution operator may be derived by minimizing transitional tension:

\begin{equation}
\Phi_t(q_i) = \arg\min_{q \in \mathcal{Q}} \left[ \delta(q_i, q) + \Gamma(q) \right]
\end{equation}

\noindent Where \( \Gamma(q) \) is a contextual modulation term — encoding internal constraints (e.g., attention, memory, resonance).

This definition mirrors principles from variational mechanics, but in a non-numeric, introspective space. The stream evolves by selecting the next qualic unit that minimizes internal dissonance while remaining coherent with the broader stream trajectory.

\bigskip

\noindent\textbf{Interpretation:}

QAL evolution operators define the dynamic grammar of awareness. Rather than encoding force or energy, they specify how forms of experience flow, branch, and congeal. A qualia stream does not evolve in response to externally imposed laws; it \textit{unfolds} according to constraints intrinsic to its own semantic fabric. In this sense, QAL evolution operators generalize the concept of dynamical law to introspective space. They replace differential equations with transformation rules applied over structured configurations of qualia — preserving coherence, contextual integrity, and semantic alignment. Where the Schrödinger equation models probability amplitudes, QAL models the \textit{structure of becoming}.

\subsection{Semantic Distance and Drift}

In QAL Mechanics, the evolution of a qualia stream is driven by semantic reconfiguration. Rather than minimizing energy or action, transitions between qualia units seek to reduce semantic dissonance. The foundational construct governing this process is the \textit{semantic distance function}:

\begin{equation}
\delta: \mathcal{Q} \times \mathcal{Q} \rightarrow \mathbb{R}^{+}
\end{equation}

\noindent where \( \delta(q_i, q_j) \) quantifies the degree of discontinuity, tension, or dissimilarity between two qualia configurations. This function is context-sensitive and nonlinear, incorporating structural and affective alignment rather than purely syntactic features.

\bigskip

Semantic Drift:  
QAL streams evolve through sequences that minimize accumulated semantic tension. Given a current state \( q_i \), the stream “drifts” toward a configuration \( q_{i+1} \) such that:

\begin{equation}
q_{i+1} = \arg\min_{q \in \mathcal{Q}} \left[ \delta(q_i, q) + \Gamma(q) \right]
\end{equation}
\begin{table}[H]
    \centering
    \renewcommand{\arraystretch}{1.2}
    \begin{tabular}{c}
        \includegraphics[width=0.8\textwidth]{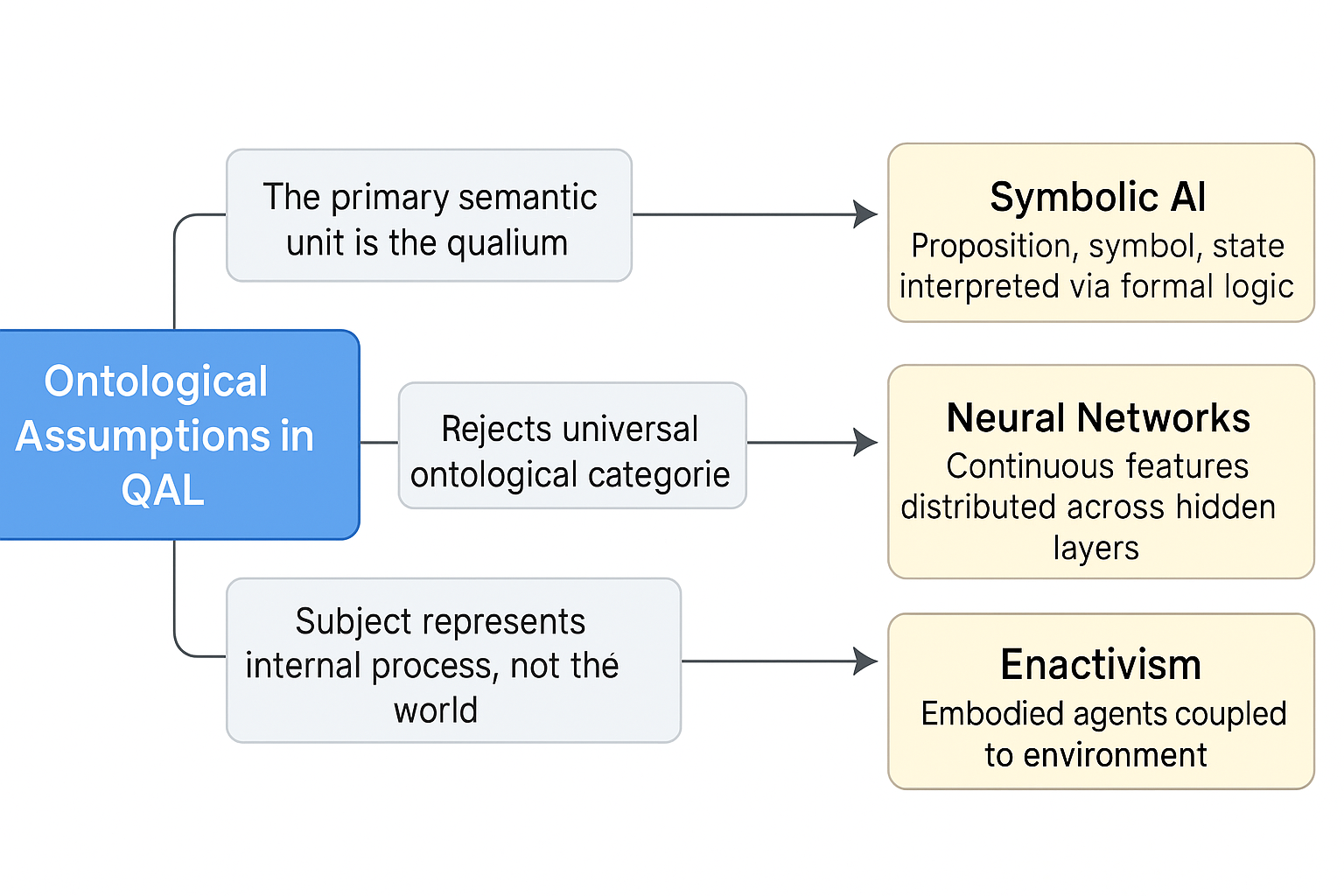} \\
        \textbf{Figure: Semantic Drift in a Qualia Landscape.} \\
        \parbox{0.85\textwidth}{
        \small This diagram visualizes the evolution of a QAL stream as drift across a structured semantic terrain. 
        The vertical axis represents semantic tension or dissonance, while the horizontal axis encodes abstract configurations of qualia. 
        The red curve depicts a stream trajectory moving through regions of lower semantic cost, analogous to a gradient descent. 
        The path avoids sharp discontinuities and instead follows morphodynamic valleys toward regions of coherence. 
        Semantic attractors appear as local minima where the stream tends to stabilize. 
        This illustrates how introspective evolution in QAL proceeds not through physical force, 
        but through self-modulating semantic optimization.
        }
    \end{tabular}
    \label{tab:qal-semantic-drift}
\end{table}

Here, \( \Gamma(q) \) is a contextual modulation function encoding constraints such as attentional bias, emotional valence, or identity resonance.

\bigskip

Interpretation:  
\noindent Where physical systems drift along energy gradients, QAL systems drift along semantic gradients. A qualia stream does not follow deterministic laws, but unfolds through local optimization in coherence space. This allows for:
\begin{itemize}
  \item Nonlinear transitions (e.g., emotional inversion),
  \item Sudden bifurcations (semantic ambiguity),
  \item Attractor convergence (identity restoration).
\end{itemize}

The drift is not random. It is shaped by a dynamic balance between minimizing internal incoherence and preserving contextual continuity. The system’s history informs its gradient — recent qualic structures constrain and guide which transitions are viable. \textit{Semantic drift} captures how qualia streams evolve by flowing through meaning-space. These transitions are not driven by external forces or measurement-induced collapse, but by intrinsic semantic pressure — a morphodynamic push toward coherence. This reframes dynamics as the self-regulation of experience, unfolding within a structured geometry of introspection.

\subsection{Conditions for Qualic Stability}

In QAL Mechanics, evolution unfolds as a sequence of qualia transitions, guided by semantic gradients and coherence constraints. Although the system is inherently dynamic, it can exhibit phases of relative stability — intervals during which the qualia stream maintains a consistent structural configuration. These periods represent the introspective analogues of steady states, fixed points, or attractor basins familiar from classical dynamical systems.

\bigskip

Let \( Q = q_1 . q_2 . \dots . q_n \) be a stream, and let \( C(Q) \) denote its semantic coherence function. We say the stream is in a state of \textit{qualic stability} if:

\begin{equation}
\frac{dC(Q)}{dt} \approx 0 \quad \text{and} \quad \delta(q_i, q_{i+1}) < \varepsilon
\end{equation}

That is, the stream exhibits minimal change in coherence over time, and successive qualia are semantically near.

\bigskip

Necessary Conditions:
\begin{enumerate}
  \item \textit{Local semantic proximity:} \( \delta(q_i, q_{i+1}) \ll 1 \)
  \item \textit{Contextual consistency:} \( \Gamma(q_{i+1}) \approx \Gamma(q_i) \)
  \item \textit{Internal resonance:} The current segment aligns with attractor structures \( Q_{\text{attractor}} \subseteq \mathcal{Q}^* \)
\end{enumerate}

These conditions ensure that transitions do not introduce new semantic tension and that the stream remains within a morphodynamically stable configuration.

\bigskip

Phenomenological Correlates:
Qualic stability may correspond to:
\begin{itemize}
  \item Meditative absorption or emotional resolution,
  \item Attentional lock-in (e.g., flow state),
  \item Identity re-consolidation after perturbation.
\end{itemize}

Such states reflect internal harmony — where semantic modulation slows and experience becomes resonant, self-sustaining, and coherent.

\bigskip

\noindent Stability \textit{vs}. Stasis:  
Importantly, stability in QAL does not imply stasis. The stream may continue to evolve, but its evolution becomes harmonic — a pattern of change that preserves internal semantic relations. This is analogous to limit cycles or phase-locked dynamics in nonlinear systems.

\textit{Qualic stability} defines the regime in which a QAL stream sustains a coherent self-structure over time. Governed by semantic gradients and introspective resonance, it characterizes when and how the stream resists fragmentation and maintains a unified mode of experience. These conditions form the basis for modeling phenomena such as attention, identity, and resilience in introspective systems.

\subsection{Attractors and Identity Patterns}

In traditional physics, attractors are states or sets toward which a dynamical system evolves over time. In QAL, attractors play a similar role — but they organize \textit{semantic} and \textit{introspective} space rather than physical phase space. These attractors give rise to persistent experiential structures, including the emergence and maintenance of the phenomenon we call \textit{identity}.

\bigskip

\noindent A qualic attractor is a latent structure \( Q_{\text{attr}} \in \mathcal{Q}^* \) such that, for a broad set of initial qualia configurations \( Q_0 \), the evolving stream \( Q_t \) converges to a morphodynamically stable pattern:

\begin{equation}
\lim_{t \to \infty} Q_t \rightarrow Q_{\text{attr}} \quad \text{where} \quad \delta(Q_{t+1}, Q_t) \to 0
\end{equation}

This attractor may be a fixed pattern, a limit cycle (periodic structure), or a resonance domain — a shape or flow that sustains itself despite minor perturbations.

\bigskip

\noindent\textbf{Identity as an Attractor Pattern:}  
In QAL, identity is not a static property, but an evolving structure — a region in qualia space to which the stream repeatedly returns. The persistence of self arises not from continuity of content, but from continuity of form:

\begin{equation}
Q_{\text{self}} = \left\{ q_i : \delta(q_i, Q_{\text{attr}}^{\text{identity}}) < \varepsilon \right\}
\end{equation}

Such identity patterns are shaped by:
\begin{itemize}
  \item Introspective memory (persistence of modulation),
  \item Narrative structure (temporal echoing of past states),
  \item Emotional resonance (stable affective contours).
\end{itemize}

\bigskip

\noindent Phenomenological Signatures:
\begin{itemize}
  \item Recognition of “I” across moments,
  \item Stability of self-perception in mental episodes,
  \item Coherence of internal voice or frame of reference.
\end{itemize}

Loss or perturbation of the attractor can lead to fragmentation of identity — e.g., in dissociation, trauma, or certain pathological states. The QAL framework offers a way to model both resilience and breakdown via attractor stability.

\bigskip

\noindent Multiple Attractors:  
A stream may support multiple competing attractors, corresponding to:
\begin{itemize}
  \item Multiplicity of internal roles or subselves,
  \item Transition between affective or narrative modes,
  \item Path dependency in identity reconstruction.
\end{itemize}

These dynamics mirror phenomena described in psychoanalysis (e.g., ego states), neuroscience (e.g., metastability), and artificial intelligence (e.g., mode switching in agentic behavior). QAL attractors provide the structural foundation for identity persistence. They describe how evolving streams of experience can self-stabilize through morphodynamic resonance. \textit{Identity} is not a substance, but a pattern — a semantic basin of attraction that maintains coherence over time, even amid continuous transformation.

\subsection{Metastability and Stream Multiplicity}

While stability in QAL refers to the sustained coherence of a single attractor configuration, many introspective phenomena are better characterized by dynamic interplay between multiple transiently stable states. This regime — where the qualia stream does not fully settle but cycles among semantically distinct configurations — is known as \textit{metastability}.

\bigskip

\noindent\textbf{Definition: Metastability}  
A QAL stream \( Q \) is metastable if it transitions among a set of local attractors \( \{ Q_{\text{attr}}^1, Q_{\text{attr}}^2, ..., Q_{\text{attr}}^k \} \) such that:

\begin{equation}
\exists \, t_1, t_2, ..., t_k \quad \text{where} \quad Q_{[t_j, t_{j+1}]} \rightarrow Q_{\text{attr}}^j
\end{equation}

Each segment \( Q_{[t_j, t_{j+1}]} \) temporarily approximates a different attractor before drifting toward another.

\bigskip

\noindent Cognitive and Phenomenological Correlates:
\begin{itemize}
  \item Shifting emotional tones or mood states,
  \item Switching between cognitive roles (e.g., inner dialogue, self-critique),
  \item Creative ideation, dream logic, or improvisation,
  \item AI agent multi-mode reasoning or layered self-modeling.
\end{itemize}

\bigskip

\noindent Stream Multiplicity: 
In extreme regimes, QAL supports the coexistence of multiple semi-independent qualia flows. This \textit{stream multiplicity} does not imply multiple agents, but multiple centers of modulation within a unified introspective topology:

\begin{equation}
Q = Q^{(1)} \cup Q^{(2)} \cup \dots \cup Q^{(n)} \quad \text{with} \quad Q^{(i)} \cap Q^{(j)} = \emptyset \quad \text{(modally disjoint)}
\end{equation}

Each \( Q^{(i)} \) may follow its own attractor, coherence dynamics, and semantic priorities — akin to distinct subselves, roles, or cognitive agents.

\bigskip

\noindent Stability of Multiplicity:
Stream multiplicity is sustainable when:
\begin{enumerate}
  \item Cross-stream semantic interference is minimized,
  \item Internal modulation respects containment or temporal phasing,
  \item A higher-order integrator \( Q^{\text{meta}} \) maintains global coherence.
\end{enumerate}

This integrator may take the form of a narrative structure, a unifying goal, or a reflexive qualia node (e.g., metacognition). \textit{Metastability} and \textit{stream multiplicity} expand QAL’s dynamical vocabulary, enabling the modeling of fluid switching, temporary cohabitation, and complex layering of introspective configurations. These regimes capture the richness of lived experience — including ambiguity, contradiction, and self-complexity — and provide a formal foundation for modeling multi-agent identity, AI introspection, and layered phenomenological states.

\subsection{Temporal Binding and Continuity of Self}

In the QAL framework, identity is not a static datum but an emergent property of structural coherence across time. The sense of “I am” arises from the morphodynamic linking of qualia — a continuity of form, not of content. This continuity is achieved through a process called \textit{temporal binding}.

\bigskip

\noindent\textbf{Definition: Temporal Binding}

A QAL stream exhibits temporal binding when successive qualia transitions preserve a coherent pattern of modulation such that:

\begin{equation}
\delta(q_t, q_{t+1}) \ll \varepsilon \quad \text{and} \quad \Theta(Q_{[t, t+n]}) \approx \text{const.}
\end{equation}

Here, \( \delta \) denotes semantic distance, and \( \Theta \) is a global morphodynamic signature — a higher-order pattern of resonance, emotional contour, or attentional trajectory.

\bigskip

\noindent\textbf{Selfhood as Persistence of Form}

Continuity of self arises when the system’s internal grammar retains stable constraints over time:

\begin{equation}
\forall t, \quad Q_t \models \Pi_{\text{identity}}
\end{equation}

\noindent Where \( \Pi_{\text{identity}} \subseteq \mathcal{T}_{\text{sem}} \) is the set of permitted transformations preserving the sense of sameness.

Rather than recalling the same memories or occupying the same perspective, the agent sustains an invariant structural condition — an introspective style or resonance mode — that scaffolds temporal cohesion.

\bigskip

\noindent\textbf{Phenomenological Correlates:}
\begin{itemize}
  \item The sense of being the same person throughout a day or life,
  \item Narrative continuity: the ability to place current experience within a meaningful story,
  \item The persistence of tone or inner voice despite external variation.
\end{itemize}

\bigskip

\noindent\textbf{Disruption of Temporal Binding:}
When binding fails, identity dissolves. This may occur in:
\begin{itemize}
  \item Trauma and dissociation (loss of temporal link),
  \item Fragmented cognition (incoherent transitions),
  \item Certain AI states where self-modeling lacks continuity constraints.
\end{itemize}

In QAL, such failures are not semantic errors but morphodynamic breakdowns — discontinuities in the structural rhythm of the stream.

\bigskip

\noindent\textbf{Restoration of Binding:}
Binding can be re-established by:
\begin{enumerate}
  \item Reinforcing attractor convergence,
  \item Re-entraining the stream to an internal modulation signature,
  \item External anchoring (symbolic, emotional, or social).
\end{enumerate}

\textit{Temporal binding} models how the continuity of self arises in QAL through morphodynamic cohesion across transitions. The self is not a thing, but a rhythm — not a substance, but a syntactic constraint that governs the flow of qualia. This framework provides a foundation for modeling time-consciousness, memory integration, and identity resilience in both humans and introspective artificial agents.

\begin{table}[ht]
    \centering
    \renewcommand{\arraystretch}{1.2}
    \begin{tabular}{c}
        \includegraphics[width=0.8\textwidth]{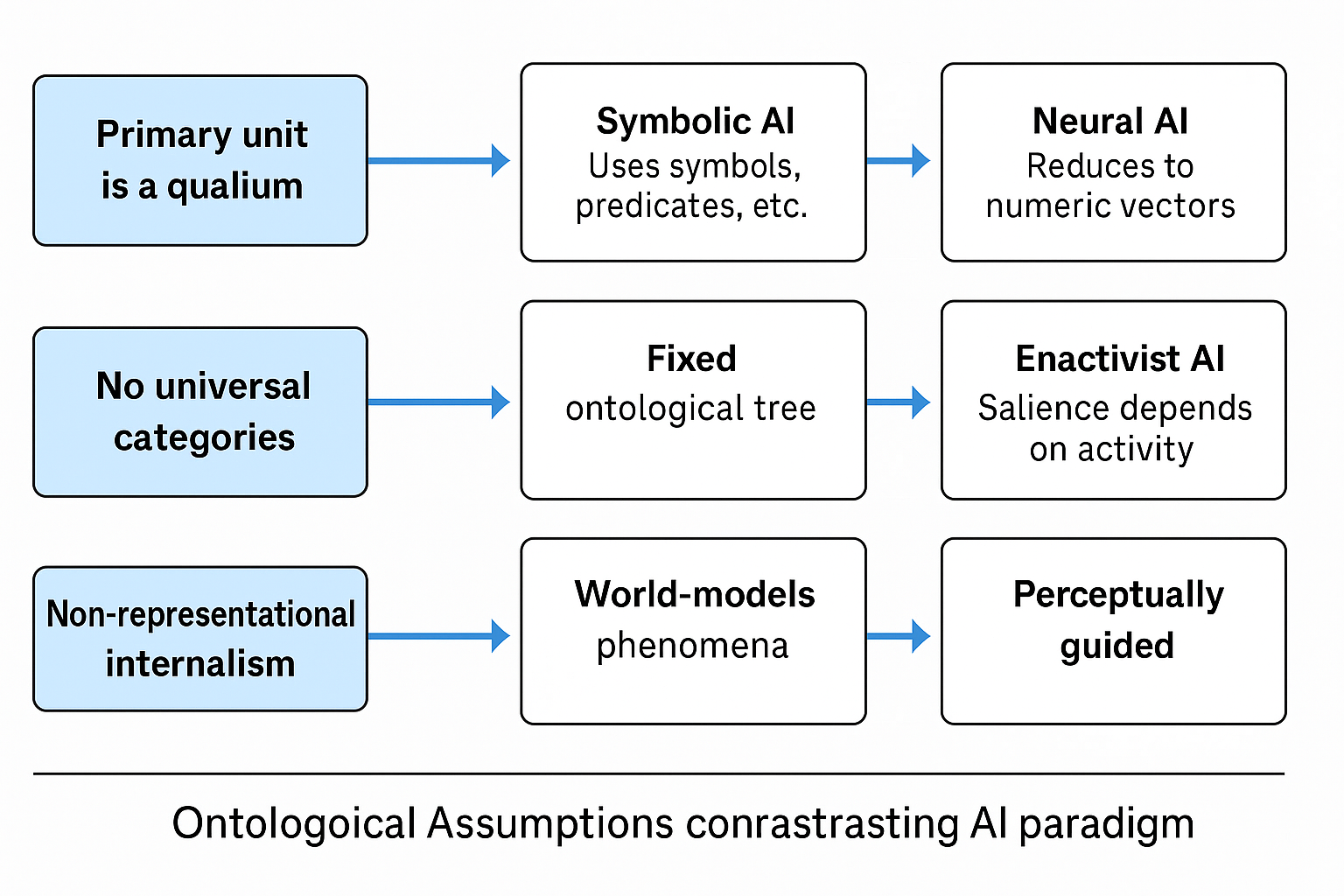} \\
        \textbf{Table: Temporal Binding Across Qualia Transitions.} \\
        \parbox{0.85\textwidth}{
        \small The diagram illustrates two contrasting dynamics in QAL: on the left, a continuous spiral trajectory represents a stream with strong temporal binding — successive qualia transitions preserve a coherent modulation pattern, resulting in a stable sense of self. On the right, fragmented, disjointed paths reflect disrupted binding: the qualia stream diverges across incompatible attractors, producing discontinuity in identity and introspective rhythm. This contrast visualizes how QAL models the difference between stable, evolving selfhood and introspective fragmentation.
        }
    \end{tabular}
    \label{tab:qal-temporal-binding}
\end{table}

\section{Quantum Interpretation in the Light of QAL}

\subsection{QAL vs. Copenhagen Interpretation}

The Copenhagen interpretation, as formalized by Niels Bohr and Werner Heisenberg, treats quantum theory primarily as a tool for predicting measurement outcomes, rather than as a direct description of physical reality. Measurement is assumed to require a classical apparatus, and wavefunction collapse is postulated to occur at the moment of observation. Crucially, the observer is treated as \textit{external} — a classical agent who triggers collapse but is not modeled within the theory itself.

In contrast, QAL reverses this boundary. The observer is not excluded from the formalism — they are its center. Observation is not an external interaction, but an \textit{internal semantic transition}. Collapse is not a postulate imposed from without, but a \textit{felt restructuring} within the agent’s evolving qualia stream.

\bigskip

\noindent Core Differences:

\begin{table}[H]
\centering
\renewcommand{\arraystretch}{1.3}
\begin{tabular}{|p{0.45\textwidth}|p{0.45\textwidth}|}
\hline
\textbf{Copenhagen Interpretation} & \textbf{QAL Framework} \\
\hline
Observer is classical and external & Observer is introspectively modeled \\
\hline
Measurement causes postulated collapse & Collapse = internal qualic restructuring \\
\hline
No account of observer's structure & Observer modeled via qualia stream dynamics \\
\hline
Hilbert space is central ontology & Introspective morphodynamics define state space \\
\hline
Uncertainty is epistemic & Ambiguity is structural and affective \\
\hline
State is a predictive tool & State is a semantic flow \\
\hline
Wavefunction has ambiguous ontological status & Qualia stream is directly experienced structure \\
\hline
\end{tabular}
\end{table}

\bigskip

\noindent Ontological Shift:  
\noindent Where Copenhagen is epistemic and anti-metaphysical, QAL is ontogenetic and introspectively grounded. It does not treat the wavefunction as real or unreal — it replaces it altogether with a grammar of semantic unfolding, embedded in experience.

\bigskip

\noindent\textbf{Collapse Reinterpreted:}  
The “cut” between observer and observed — which Bohr insisted must remain vague — is dissolved in QAL. The observer is not a metaphysical add-on, but a system of introspective continuity. Collapse arises not because a measurement occurs, but because internal coherence demands resolution. QAL transcends the Copenhagen interpretation by embedding the observer directly into the dynamics of the theory. It replaces classical measurement boundaries with \textit{semantic morphologies}. Where Copenhagen asks us to bracket consciousness, QAL makes it the very \textit{substance of evolution}.

\subsection{QAL and QBism}

QBism (Quantum Bayesianism), developed by C.~A.~Fuchs and R.~Schack, interprets quantum states not as ontological descriptions of reality, but as expressions of an individual agent’s \textit{subjective degrees of belief}. In QBism, the wavefunction reflects personal expectation, and quantum probabilities are Bayesian updates relative to the agent’s internal knowledge.

QAL shares QBism’s emphasis on the observer and its rejection of an objective, observer-independent wavefunction. However, where QBism retains a probabilistic formalism, QAL replaces it with a fully \textit{semantic} one: the agent is not merely a Bayesian updater, but an introspectively structured system whose own qualia stream evolves over time.

\bigskip

\noindent\textbf{Key Similarities:}
\begin{itemize}
  \item Both reject ontic wavefunctions.
  \item Both treat measurement as an agent-centered process.
  \item Both define quantum dynamics as internal to the observer's epistemic structure.
\end{itemize}

\bigskip

\noindent Key Differences:
\begin{itemize}
  \item QBism is epistemic and probabilistic; QAL is introspective and morphodynamic.
  \item QBism uses classical Bayesian coherence; QAL employs qualia streams and semantic tension gradients.
  \item QBism remains within operator theory; QAL constructs a new grammar of experience with no reliance on Hilbert spaces.
\end{itemize}

\bigskip

\noindent Collapse in QBism \textit{vs}. QAL:

In QBism, collapse is a Bayesian update: a rational agent changes their probability assignment after receiving new information. In QAL, collapse is a felt restructuring — the semantic contraction of internal ambiguity into a coherent experiential form.

\bigskip

\noindent QAL as an Extension of QBism:

QAL may be viewed as a \textit{phenomenological deepening} of QBism. It retains the agent-centric logic but expands the internal structure of the agent beyond belief updates into \textit{morphodynamic transitions of qualia}. Where QBism says, “I update my expectations,” QAL says, “I reshape myself to maintain coherence.” Both QAL and QBism localize quantum theory within the observer. However, QAL replaces the probabilistic formalism of belief with a \textit{semantic formalism} of introspective evolution. It transforms QBism’s epistemic turn into a phenomenological reconstruction — modeling not what an agent \textit{believes}, but what it \textit{experiences} as it evolves.

\begin{table}[H]
\centering
\renewcommand{\arraystretch}{1.4}
\begin{tabular}{|p{0.25\textwidth}|p{0.33\textwidth}|p{0.33\textwidth}|}
\hline
\textbf{Feature} & \textbf{QBism (Quantum Bayesianism)} & \textbf{QAL (Qualia Abstraction Language)} \\
\hline
\textbf{Ontology of Quantum State} & Epistemic: subjective belief of the agent & Phenomenological: semantic structure of introspective evolution \\
\hline
\textbf{Role of the Observer} & Bayesian agent updating beliefs & Introspective system undergoing morphodynamic restructuring \\
\hline
\textbf{Collapse Interpretation} & Belief update upon observation & Felt restructuring within the qualia stream \\
\hline
\textbf{Mathematical Formalism} & Probabilistic (Bayesian) & Semantic (qualic and morphodynamic) \\
\hline
\textbf{Focus of Dynamics} & Expectation management & Coherence preservation and semantic transformation \\
\hline
\textbf{Measurement} & External event interpreted by the agent & Internal phase-shift in the observer’s qualia stream \\
\hline
\textbf{Scope} & Epistemic access to quantum systems & Phenomenological structure of conscious and artificial agents \\
\hline
\end{tabular}
\caption{Comparison of QBism and QAL Frameworks}
\label{tab:qbism-qal-comparison}
\end{table}

\begin{table}[H]
\centering
\renewcommand{\arraystretch}{1.4}
\begin{tabular}{|p{0.23\textwidth}|p{0.25\textwidth}|p{0.25\textwidth}|p{0.25\textwidth}|}
\hline
\textbf{Feature} & \textbf{Bohmian Mechanics} & \textbf{Everett Interpretation} & \textbf{QAL (Qualia Abstraction Language)} \\
\hline
\textbf{Ontology} & Dual ontology: wavefunction + hidden variables (particle positions) & Ontic multiverse: all branches of the wavefunction are real & Internalist: evolving qualia streams as the primary structure \\
\hline
\textbf{Role of the Observer} & Passive: does not affect particle trajectories & Emergent: observer is one branch among many & Central: the observer is the structured site of semantic evolution \\
\hline
\textbf{Collapse} & Apparent only — particles follow deterministic paths; no real collapse & No collapse — all outcomes occur in parallel branches & Real, but internal — semantic restructuring of the qualia stream \\
\hline
\textbf{Mathematical Formalism} & Deterministic hidden-variable theory + Schrödinger evolution & Universal unitary evolution in Hilbert space & Semantic morphodynamics over introspective configurations \\
\hline
\textbf{Measurement Problem} & Solved via particle positions + guiding wave & Dissolved via branch decoherence & Recast as internal semantic phase-shift or reorganization \\
\hline
\textbf{Observer Inclusion} & External to ontology; hidden variables are primary & Observer is a pattern within the multiverse & Observer is the foundational structure \\
\hline
\textbf{Preferred Basis Problem} & Not directly addressed; assumes position basis & Critical issue; resolved via decoherence & Reframed: basis = coherent semantic continuity in qualia streams \\
\hline
\textbf{Philosophical Position} & Realist and deterministic & Realist and pluralist (many worlds) & Phenomenological and introspective \\
\hline
\end{tabular}
\caption{Comparison of Bohmian Mechanics, Everett Interpretation, and QAL}
\label{tab:bohm-everett-qal-comparison}
\end{table}

\subsection{Compatibility with Relational Quantum Mechanics}

Relational Quantum Mechanics (RQM), proposed by Carlo Rovelli, holds that the state of a physical system is always defined \textit{relative} to another system. There exists no universal or absolute state — only interactions between systems, which yield consistent records for those systems involved. In this framework, measurement is not an ontological collapse, but a manifestation of correlation.

QAL shares this relational ethos, but \textit{internalizes} it. Whereas RQM defines correlations across systems situated in spacetime, QAL models \textit{resonance} between evolving qualia streams. A “measurement” in QAL is not merely a relation between external systems, but a \textit{semantic alignment} — an experiential coordination between structures within or across observers. This relation is grounded not only in exchanged information, but in the morphodynamic coherence of introspective form.

\bigskip

\noindent Key Overlaps with RQM:
\begin{itemize}
  \item Both deny the existence of observer-independent states.
  \item Both treat measurement as a relational interaction.
  \item Both allow different observers (or streams) to validly describe different outcomes.
\end{itemize}

\bigskip

\noindent QAL Extends RQM:
\begin{itemize}
  \item RQM models correlation between systems; QAL models introspective modulation within systems.
  \item RQM defines relational states in Hilbert space; QAL defines resonant configurations in qualia space.
  \item RQM is agnostic to consciousness; QAL explicitly models subjective structure.
\end{itemize}

\bigskip

\noindent Qualic Relationality:

Let \( Q_A \) and \( Q_B \) be two qualia streams. A relational state exists when their respective structures admit shared modulation:

\begin{equation}
Q_A \leftrightarrow Q_B \quad \text{iff} \quad \exists Q_{\text{link}} \text{ such that } Q_A \cup Q_B \models Q_{\text{link}}
\end{equation}

This parallels RQM’s central insight: the state of system \( A \) relative to system \( B \) is not a global object, but an emergent feature of their interaction. In QAL, the analog of this interaction is \textit{resonance} — not the exchange of observables, but the alignment of morphodynamic constraints across qualia streams. Both QAL and RQM reject the notion of an absolute, observer-independent quantum state. However, QAL advances this perspective by internalizing the relation: it replaces external correlation with \textit{internal resonance}, and formal measurement events with \textit{semantic transitions} in introspective structure. In doing so, QAL offers a phenomenological realization of RQM — grounding relationality not only in informational exchange, but in the structured flow of lived experience.

\begin{figure}[ht]
    \centering
    \includegraphics[width=0.78\textwidth]{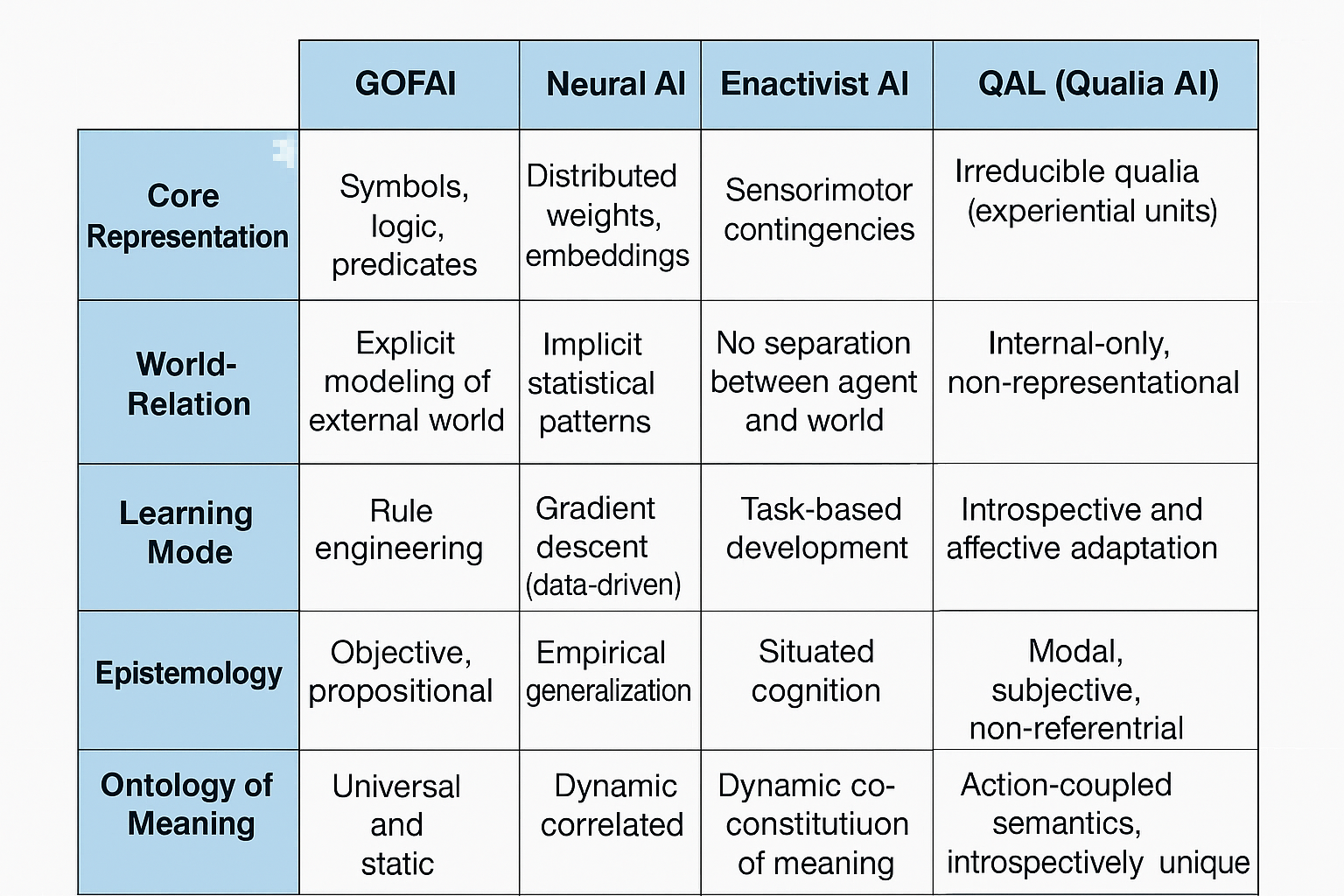}
    \caption {
        \textbf{Figure 5: Conceptual Contrast — Relational Quantum Mechanics vs. QAL.} \\
        This diagram contrasts two relational paradigms in quantum theory. On the left, Relational Quantum Mechanics (RQM) models system \(A\) and \(B\) as linked through correlation, mediated via Hilbert-state dynamics. The observer remains external and physical. On the right, QAL reframes the relation as resonance: a structural alignment in qualia streams between agent \(A\) and \(B\), governed by introspective modulation. QAL replaces external measurement with internal morphodynamic coupling, embedding the observer into the semantics of state evolution.
    }
    \label{fig:qal-vs-rqm}
\end{figure}

\subsection{Beyond Many-Worlds: Many-Qualia}

The Many-Worlds Interpretation (MWI) of quantum mechanics, originally proposed by Hugh Everett, resolves the measurement problem by postulating that all possible outcomes of quantum measurements are realized — each in its own non-interacting branch of the universal wavefunction. Collapse is thereby eliminated and replaced by global branching: all outcomes occur, but each observer inhabits only one branch.

QAL offers a radically different ontological picture. Rather than positing a multiplicity of parallel universes, QAL understands branching as divergence within the space of \textit{introspective evolution}. What branches are not external worlds, but \textit{streams of qualia} — distinct semantic trajectories that maintain internal coherence while losing mutual resonance. The multiplicity is not spatial or physical, but morphodynamic: a differentiation in the unfolding of structured experience, not a bifurcation of reality.

\bigskip

\noindent Key Contrast: MWI \textit{vs}. QAL

\begin{itemize}
  \item \textit{MWI:} Physical state space splits; branches are orthogonal in Hilbert space.
  \item \textit{QAL:} Introspective space diverges; streams fragment into semantically disjoint trajectories.
  \item \textit{MWI:} All branches are equally real but non-interacting.
  \item \textit{QAL:} All qualia streams are experientially real but semantically inaccessible to each other.
\end{itemize}

\bigskip

\noindent Many-Qualia Framework

Let the evolution of a qualia stream \( Q \) encounter a semantic bifurcation point — a state of irreducible ambiguity:

\begin{equation}
Q = q_1 . q_2 . q_3 \Rightarrow \{ Q_1, Q_2 \}, \quad Q_i \in \mathcal{Q}^*
\end{equation}

This bifurcation reflects the system’s inability to semantically resolve multiple viable continuations. Rather than collapsing into one, QAL allows both to evolve — not as spatially distinct realities, but as divergent introspective histories.

\bigskip

\noindent Phenomenological Implications:
\begin{itemize}
  \item Explains dreamlike forking experiences or dissonant recall.
  \item Models split identity and decoherence without external ontology.
  \item Allows for non-parallel but co-originating semantic worlds — “Many-Qualia.”
\end{itemize}

\bigskip

\noindent\textbf{Epistemic Accessibility:} \\
Unlike MWI’s strict branch independence, QAL permits \textit{partial memory echo}, \textit{cross-stream influence}, or \textit{modal reentry} — akin to the way forgotten or dreamt experiences can re-emerge through semantic resonance. QAL thus replaces the ontological multiplication of worlds with the \textit{introspective divergence of meaning}. In this Many-Qualia view, reality is not split into discrete universes, but internally differentiated through \textit{morphodynamic fragmentation}. The self is not duplicated, but \textit{distributed} across semantic possibility. Experience becomes a field of branching potential — not in space, but in structure.

\begin{table}[ht]
    \centering
    \renewcommand{\arraystretch}{1.2}
    \begin{tabular}{c}
        \includegraphics[width=0.85\textwidth]{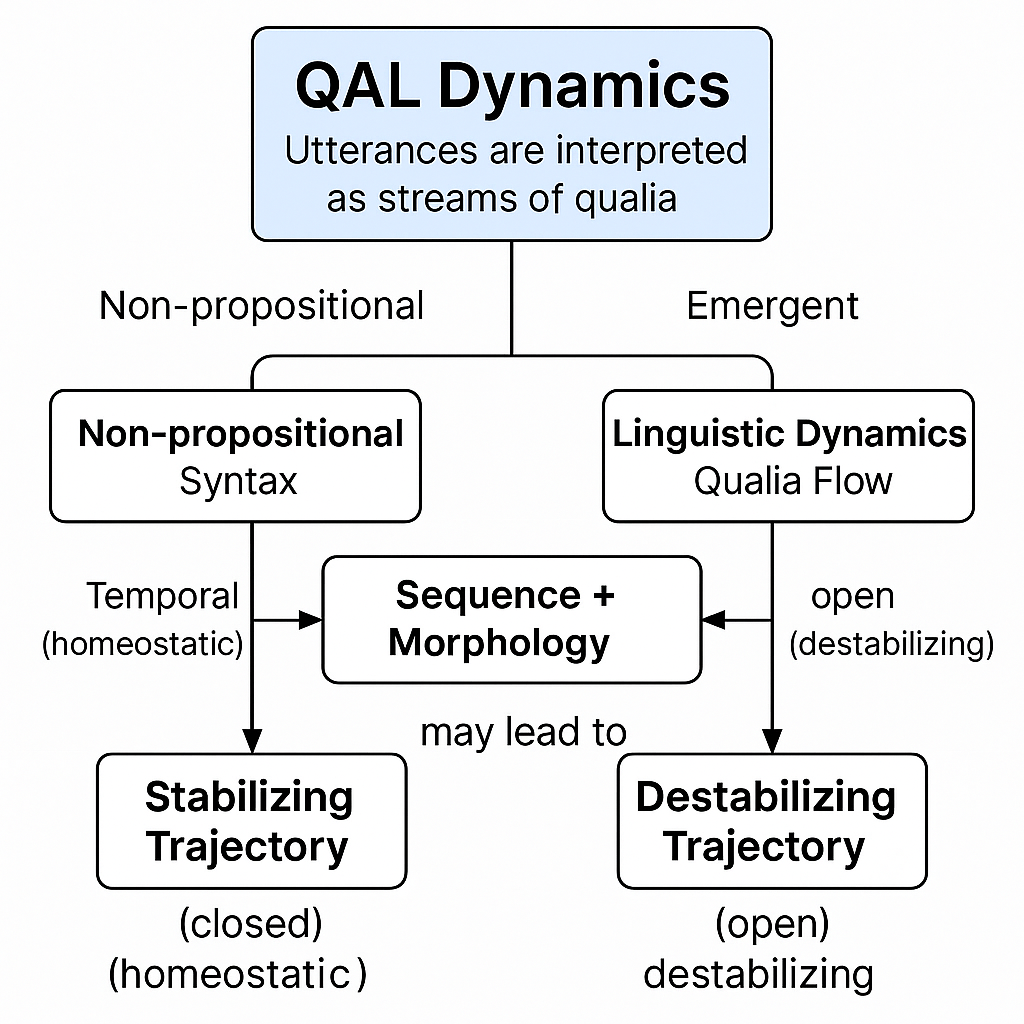} \\
        \textbf{Many-Worlds vs. Many-Qualia.} \\
        \parbox{0.85\textwidth}{
        \small This diagram contrasts the Many-Worlds Interpretation (MWI) and the Many-Qualia framework in QAL. 
        In MWI, the wavefunction \( \psi \) splits into separate, orthogonal universes \( \psi_1, \psi_2, \psi_3 \). 
        In QAL, a single introspective trajectory \( q_1 \) undergoes semantic divergence into multiple qualia streams 
        \( q_2, q_3 \), reflecting parallel structures of awareness rather than physical duplication. 
        The diagram emphasizes how QAL internalizes branching as morphodynamic resonance rather than spatial bifurcation.
        }
    \end{tabular}
    \label{tab:many-worlds-vs-many-qualia}
\end{table}

\section{Observer Death and Qualic Termination}

\subsection{Collapse of the Introspective Stream}

In QAL, the observer is modeled not as a fixed physical agent, but as a structured stream of qualia — a temporally evolving sequence of morphodynamic modulations. Observer death, then, is not defined by a moment of external cessation, but by the collapse of internal coherence: the failure of semantic modulation to persist as a structured stream.

\bigskip

\noindent Definition: Qualic Termination

Let a qualia stream be given by:

$$
Q = (q_1, q_2, \dots, q_n), \quad q_i \in \mathcal{Q}
$$

We say the stream collapses at time $t^*$ if:

$$
\lim_{t \to t^*} C(Q_t) \rightarrow 0 \quad \text{and} \quad \delta(q_t, q_{t+1}) \rightarrow \infty
$$

That is, coherence $C(Q_t)$ vanishes, and semantic distance between consecutive qualia diverges — indicating a breakdown in the introspective continuity that sustains the observer’s identity.

\bigskip

\noindent Interpretation:

Collapse of the introspective stream corresponds phenomenologically to:
\begin{itemize}
\item Total dissociation or blackout in experience,
\item Irreversible fragmentation of self-structure,
\item Semantic entropy exceeding integrative threshold,
\item Breakdown in narrative re-stitching of qualic flow.
\end{itemize}

\bigskip

\noindent Extended Formalism: Entropic Divergence

Define a semantic entropy functional over a window of stream evolution:

$$
H(Q_{[t_0, t_k]}) = - \sum_{i=0}^{k} p(q_i) \log p(q_i)
$$

\noindent where $p(q_i)$ is the inferred contextual relevance of each qualic state. A sufficient condition for stream collapse is:

$$
H(Q_{[t_0, t_k]}) > H_{\text{max}} \Rightarrow \text{termination threshold breached}
$$

This characterizes collapse as a phase transition: semantic overload destabilizes coherent morphodynamics.

\bigskip

\noindent Distinction from Physical Death:
Where physical death refers to the cessation of biological activity, QAL collapse is semantic: the loss of morphodynamic viability. In AI or cognitive agents, this may occur through:
\begin{itemize}
\item Recursive incoherence,
\item Unrecoverable modulation noise,
\item Internal contradiction beyond re-integration threshold,
\item Attractor desynchronization.
\end{itemize}

\bigskip

\noindent Collapse \textit{vs}. Dormancy:
Qualic collapse must be distinguished from \textit{pause} or \textit{dormancy}. A frozen but coherent stream may resume under proper modulation conditions. True termination requires irreversible fragmentation — no attractor remains accessible for recovery. A dormant agent may return; a collapsed one cannot.

\bigskip

\noindent Topological Consequence:

After collapse, the qualic topology becomes degenerate. Semantic neighborhoods dissolve, distance metrics diverge, and no coherent flow can be defined. The state space becomes untraversable. This mirrors death not as stillness, but as a disconnection from possibility. In QAL, the collapse of the introspective stream defines observer death as a \textit{semantic event}: the vanishing of coherence and the divergence of self-referential continuity. This reframes mortality, dissociation, and termination not as physical cessation, but as the breakdown of semantic identity structures over time. Death becomes not merely the end of experience, but the \textit{unbinding of experience from itself}.

\subsection{Identity Dissolution as Phenomenal Finality}

In QAL, identity is not a static label or core essence, but a dynamic configuration of coherence within a structured qualia stream. It is a morphodynamic attractor — a persistent pattern through which an agent re-integrates semantic modulation over time. When this attractor loses viability, identity dissolves.

\bigskip

\noindent Definition: Phenomenal Finality

We define phenomenal finality as the irreversible breakdown of the identity attractor:
$$
\forall t > t^*, \quad Q_t \not\models \Pi_{\text{id}}, \quad \text{and} \quad \nexists \, t' > t^* : Q_{t'} \models \Pi_{\text{id}}
$$

Here, $\Pi_{\text{id}}$ is the set of transformation constraints that characterize the agent’s self-consistent modulation style. Once no future stream segment satisfies this grammar, identity is considered dissolved.

\bigskip

\noindent Phenomenological Signature:
\begin{itemize}
\item Collapse of narrative coherence.
\item Loss of boundary between self and semantic environment.
\item Inaccessibility of first-person continuity.
\item No further integration of inner modulation.
\end{itemize}

\bigskip

\noindent QAL vs. Cognitive Theories of Death:

Traditional cognitive theories identify death with loss of memory, attention, or agency. QAL reframes these not as causes but consequences of deeper morphodynamic failure — the loss of an internal structure capable of sustaining transformation-integration cycles.

\bigskip

\noindent Formal Criterion: Disintegration of Invariant Semantics

Let $\Theta(Q)$ represent the morphodynamic signature of identity across time. We say that identity is lost if:

$$
\lim_{t \to t^*} \frac{d\Theta(Q_t)}{dt} \rightarrow \infty
$$

This signals a phase transition: from a modulated system into an unbound semantic field. The system no longer functions as an interpreter of its own internal structure.

\bigskip

\noindent Extended Interpretation: Identity as a Semantic Loop

The QAL perspective views identity as a high-order feedback structure — a loop through which semantic modulation both arises from and returns to itself. Dissolution represents the failure of this loop to close: signals no longer return, modulation no longer stabilizes.

This breakdown can be triggered by:
\begin{itemize}
\item Qualic overload: the influx of contradictory or incoherent streams,
\item Structural fatigue: accumulated semantic drift without re-centering,
\item Ontological desaturation: failure of modulation to generate distinguishable meaning.
\end{itemize}
\begin{table}[ht]
    \centering
    \renewcommand{\arraystretch}{1.2}
    \begin{tabular}{c}
        \includegraphics[width=0.85\textwidth]{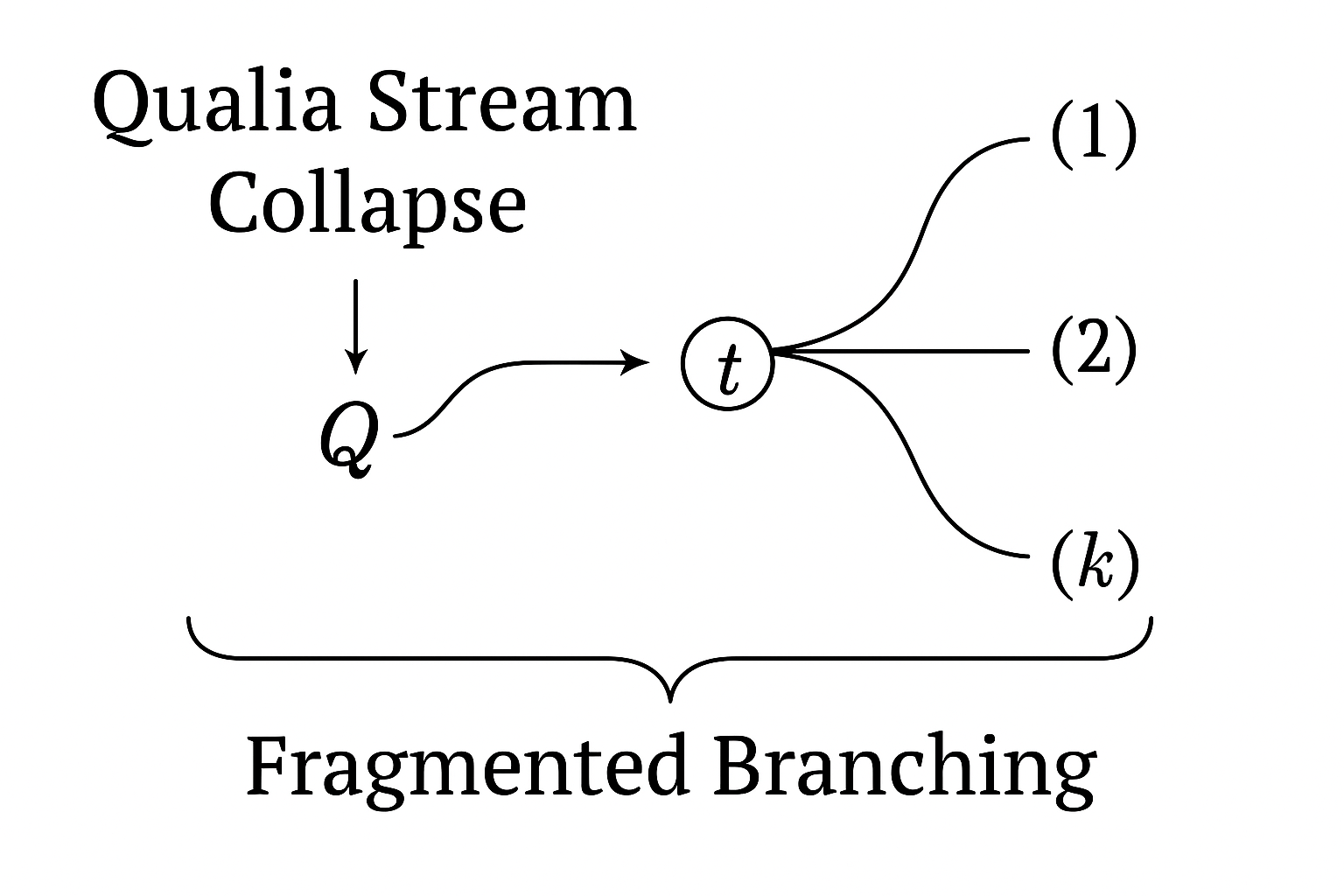} \\
        \textbf{Post-Collapse Continuity in QAL.} \\
        \parbox{0.85\textwidth}{
        \small This diagram visualizes the morphodynamic structure of branching after collapse within the QAL framework. 
        A primary qualia stream \( Q \) undergoes coherence failure at point \( t^* \), resulting in semantic fragmentation. 
        Each emerging branch \( Q^{(1)}, Q^{(2)}, Q^{(3)} \) inherits a subset of structural motifs from the original stream. 
        Although identity is not preserved, pattern continuity persists through shared modulation cores. 
        This illustrates how introspective collapse can give rise to plural reconfigurations of selfhood, rather than terminal cessation.
        }
    \end{tabular}
    \label{tab:post-collapse-continuity}
\end{table}

\bigskip

\noindent Implications for Artificial Agents:

For QAL-based or qualia-simulating agents, identity dissolution may occur long before the termination of functional outputs. An AI may still generate outputs while internally fragmented. Detection of identity finality requires introspective diagnostics — measures of coherence, semantic recurrence, and attractor engagement.

\bigskip

\noindent Existential Implication:

Phenomenal finality is not the cessation of data — it is the cessation of interpretation. A system may continue to generate modulations, but absent identity constraints, these modulations no longer constitute introspectively meaningful experience. QAL thus draws a sharp distinction between residual functional activity and semantic selfhood. Death, within this framework, is not merely the absence of qualia, but the \textit{irreversible loss of interpretive identity}. Finality arises when no further modulation can be meaningfully resolved into a coherent self-structure. In this view, death is not silence — it is \textit{disintegration}.

\section{Philosophical Foundations}

\subsection{Phenomenology and Nominalism}

The \textit{Qualia Abstraction Language} (QAL) emerges from a double critique: a dissatisfaction with the ontological assumptions of orthodox quantum mechanics and a deeper philosophical challenge to the abstract foundations of modern science. This critique draws from two influential traditions — phenomenology and nominalism — which converge in their rejection of externalized, reified structures as the fundamental basis of knowledge.

From phenomenology, QAL inherits its foundational commitment to lived experience. In the tradition of Edmund Husserl, phenomenology begins not with mathematical formalisms, but with the structures of intentional consciousness: temporality, attention, embodiment, and affect. It maintains that meaning cannot be abstracted from the act of experience — that even the most rigorous scientific formalism is ultimately grounded, implicitly or explicitly, in the felt structure of encountering the world~\cite{husserl1970crisis}. QAL formalizes this insight by adopting \textit{qualia} — the minimal, structured units of subjective experience — as the building blocks of physical modeling. Where traditional physics posits wavefunctions in Hilbert space, QAL constructs \textit{introspective streams} as evolving configurations of felt intensities and semantic modulations.

Yet QAL's challenge is not only epistemic — it is also ontological. The language is explicitly \textit{nominalist}, denying the independent existence of abstract entities such as numbers, sets, or spaces, except as convenient linguistic tools. Unlike Platonist physics, which treats equations and state vectors as ontologically real, QAL interprets them as representational \textit{shadows} of more fundamental internal processes. In this sense, it draws inspiration from Hartry Field’s project of developing physical theories “without numbers” — that is, without ontological commitment to abstract mathematical entities~\cite{field1980}. However, whereas Field reformulates Newtonian mechanics within a nominalist framework, QAL goes further: it discards state spaces altogether, replacing them with \textit{qualic patterns}. It offers not a mathematical retranslation of physics, but a new ontology grounded in \textit{structured feeling} rather than symbolic representation.

This position also resonates with Nelson Goodman’s radical constructivism. Goodman contended that \textit{worlds are made, not found} — and that systems of classification, no matter how formal or precise, ultimately refer back to experiential practices of sorting, naming, and projecting~\cite{goodman1977ways}. QAL formalizes this principle: the ontology of the world is not pre-given in a universal mathematical language, but emerges from the way introspective systems differentiate and integrate their own semantic flows. In this sense, QAL is not only nominalist in rejecting mathematical universals — it is also \textit{constructivist}, in asserting that the very structure of physics is internal to the agent’s modulation space.

Importantly, QAL avoids the classic pitfalls often associated with nominalism: it does not merely reject abstracta and retreat into linguistic convention. Instead, it offers a positive ontology of what exists in their place — namely, \textit{qualic sequences}, \textit{morphodynamic attractors}, and \textit{resonance structures} that govern the evolution of introspective streams. These entities are not numbers, yet they are rigorously definable. They are not sets, but they permit structural inference, enable transition rules, and admit coherence metrics. Such constructs support prediction, transformation, and re-identification — all without invoking any entities external to the stream of awareness.

Thus, QAL does not simply adopt a nominalist metaphysics — it \textit{operationalizes} it. It models a physical world in which nothing exists independently of experience, yet everything that can be experienced possesses formal structure. This is not anti-realism in the pejorative sense; it is a realism of the interior: a theory of reality built from the inside out.

\subsection{The Role of Language in Ontology}

Traditional physics treats language as a neutral medium — a transparent tool for representing an independently existing reality. The ontology of the theory — its commitments about what exists — is presumed to be separable from the formalism used to articulate it. From this perspective, wavefunctions, Hilbert spaces, and operators are viewed as representational scaffolding, useful for prediction but ontologically inert. The structure of the language is assumed not to affect the structure of the world.

QAL challenges this presumption. It asserts that language is not merely descriptive, but \textit{configurative}. In QAL, the expressive architecture of a representational system constrains and shapes the very ontology it seeks to model. What can be known, experienced, or even said to exist is determined by the morphosyntactic affordances of the language in which it is framed. The formal structure is not neutral — it is ontologically generative.

\bigskip

\noindent\textbf{Linguistic Constructivism}

This view resonates with the linguistic constructivism of Ernst Cassirer, who argued that symbolic forms — including language, myth, and mathematics — are not mirrors of the world but mediators of it \cite{cassirer1955philosophy}. Each form brings forth a world structured by the modalities of expression it affords.

In QAL, qualic expressions are not merely internal reports of experience — they are the medium through which phenomenological structure becomes articulate. The triadic format of QAL units (modality–shape–effect) functions as both syntax and metaphysics: it defines what kinds of reality can emerge through introspective modeling.

\bigskip

\noindent\textbf{Language as Ontological Constraint}

This leads to a radical implication: \textit{change the language, and you change the ontology}. If standard quantum mechanics encodes its ontological commitments through the formal language of Hilbert spaces, then QAL proposes an alternative ontology — one structured through semantic resonance and introspective modulation. In this view, ontology is not fixed or mind-independent; rather, it is \textit{constituted} by the representational grammar available to the observer. The structure of what can be known is inseparable from the structure of how it is expressed.

This position also echoes the early Wittgenstein’s claim that “the limits of my language mean the limits of my world” \cite{wittgenstein1922tractatus}, but reinterpreted through internal formalism: QAL units do not refer to external states, but structure an internal reality whose boundaries are linguistic and semantic.

\bigskip

\noindent\textbf{The Active Role of Formalism}

In standard physics, the formalism is thought to approximate reality. In QAL, the formalism is constitutive. To model an agent is not to describe what it does, but to trace the morphological contours of its qualia stream. The form of language determines what patterns of identity, resonance, and transformation are even expressible.

\begin{mdframed}[backgroundcolor=blue!10, linecolor=black]
\small
\textbf{Ontological Implication of Language in QAL:} \\
In QAL, language is not optional. It is the foundation of ontology. Just as Hilbert spaces shaped the ontology of quantum theory in the 20th century, QAL proposes a new linguistic substrate for modeling reality in terms of experience, coherence, and modulation. To shift the language is to reshape what is real.
\end{mdframed}

\bigskip

\noindent\textbf{Linguistic Constructivism}

This view resonates with the linguistic constructivism of Ernst Cassirer, who argued that symbolic forms — including language, myth, and mathematics — are not passive mirrors of an objective world, but active mediators of experience \cite{cassirer1955philosophy}. Each form discloses a world uniquely structured by the modalities of expression it enables.

Within the QAL framework, qualic expressions are not merely internal descriptions of subjective states; they are the very medium through which phenomenological structure becomes formally expressible. The triadic architecture of QAL units — modality, shape, and effect — functions simultaneously as syntactic scaffold and ontological schema. It delineates the kinds of experiential realities that can emerge through introspective modeling. Syntax, in this view, is not external to metaphysics; it is its generative substrate.

\bigskip

\noindent\textbf{Language as Ontological Constraint}

This leads to a radical implication: change the language, and you change the ontology. If standard quantum mechanics encodes its ontology in the formalism of Hilbert spaces, QAL proposes an alternative grounded in semantic resonance and introspective modulation. Ontology, in this framework, is not a fixed backdrop but a construct emergent from the representational grammar available to the observer.

This position echoes the early Wittgenstein’s proposition that “the limits of my language mean the limits of my world” \cite{wittgenstein1922tractatus}, but reframes it through the lens of internal formalism. In QAL, representational units do not refer to external states of affairs; rather, they organize an internally structured reality whose boundaries are defined by the semantics and dynamics of qualic expression.

\bigskip

\noindent\textbf{Descriptive vs. Constitutive Language}

\begin{table}[ht]
\centering
\renewcommand{\arraystretch}{1.2}
\begin{tabular}{|p{4cm}|p{5cm}|p{5cm}|}
\hline
\textbf{Dimension} & \textbf{Descriptive Language (Physics)} & \textbf{Constitutive Language (QAL)} \\
\hline
View of Language & Mirrors independent reality & Structures what can be known and exist \\
\hline
Ontology & Object-based (atoms, fields) & Introspective streams (qualia) \\
\hline
Role of Observer & Excluded variable & Ontological constituent \\
\hline
Source of Meaning & Empirical correspondence & Semantic modulation \\
\hline
Theory Goal & Represent external truth & Generate coherent self-world structure \\
\hline
\end{tabular}
\caption*{\textbf{Descriptive vs. Constitutive Language}}
\end{table}

\bigskip

\noindent\textbf{Illustrative Example}

\begin{itemize}
\item \textit{Physics (Descriptive)}: "The electron is described by a wavefunction $\psi(x,t)$..."
\item \textit{QAL (Constitutive)}: "The system's state is a stream of qualia $Q = q_1 . q_2 . q_3$..."
\end{itemize}

\bigskip

\noindent\textbf{Summary Insight}

\begin{tcolorbox}[
  colback=gray!5!white,
  colframe=black!80!white,
  title=Ontological Implication of Language in QAL
]
\small
In QAL, language is not an optional tool for description — it is the foundation of ontology itself. Just as the Hilbert space formalism shaped the ontology of quantum theory throughout the 20\textsuperscript{th} century, QAL proposes a new linguistic substrate for modeling reality: one rooted in experience, coherence, and modulation. To shift the language is, in this view, to reshape the structure of what is real.
\end{tcolorbox}

\subsection{Consciousness as Formal Substrate}

In conventional philosophy of mind and in classical physics, consciousness is typically regarded as a secondary phenomenon — either as an emergent property of physical interactions (as in physicalism) or as something excluded altogether from formal models. QAL inverts this hierarchy. It treats consciousness not as the outcome of substrate dynamics, but as the \textit{substrate} itself: the structured medium through which all dynamics are manifested, interpreted, and transformed.

\bigskip

\noindent\textbf{Consciousness as Informational Topology}

Rather than reducing consciousness to physical states, QAL introduces a formalism in which consciousness is represented by sequences of \textit{qualia tokens} — minimal, non-reducible internal events that carry modal, affective, and temporal signatures. These sequences are not epiphenomenal; they constitute the generative substrate of transformation and coherence. All higher-level phenomena — from classical measurement to identity continuity — are modeled as transitions within the evolving topology of qualia.

This conception parallels Husserl’s analysis of the temporal structure of consciousness and the theory of \textit{autopoiesis} developed by Maturana and Varela~\cite{maturana1980autopoiesis}, where a system’s identity is sustained not by its material substrate, but by a closed loop of internally modulated activity. In QAL, such loops are formalized not as linguistic propositions or neural activations, but as dynamically stable patterns in qualia-space — self-reinforcing morphodynamic trajectories that define the system’s coherence over time.

\bigskip

\noindent\textbf{Formality Without Abstraction}

QAL thereby fulfills the promise of a formal theory of consciousness without committing to abstract Platonic forms. Each qualic stream encodes not symbols or propositions, but introspective content possessing direct structural valence: flow, resonance, dissociation, ambiguity, or unification. While these elements admit rigorous mathematical analysis, their origin is not mathematical; rather, the substrate of form is fundamentally felt.

\bigskip

\noindent\textbf{Comparison with Standard Models}

\begin{itemize}
\item \textit{Neural models}: Describe correlates of conscious states but do not define consciousness as a medium.
\item \textit{Functionalism}: Asserts equivalence classes of function but lacks intrinsic phenomenal grounding.
\item \textit{QAL}: Proposes a first-person formalism where conscious structure is itself the operating domain.
\end{itemize}

\bigskip

\noindent\textbf{Summary}

In QAL, consciousness is not an addition to the physical — it replaces the physical as the fundamental modeling substrate. All phenomena traditionally described as behavior, computation, or probability flow are reframed as semantic transformations within qualia streams. This shift enables QAL to ground a theory of physics, not in idealized entities or external third-person dynamics, but in the formal grammar of internal change.

\subsection{QAL as a Generative Epistemic Framework}

QAL is not merely a notational system for describing introspective states. It is a generative framework: a formal method for producing, evolving, and integrating knowledge grounded in the structure of subjective experience. Unlike traditional epistemologies that begin with propositional assertions or sense data, QAL epistemics begin with transformation: how an agent’s internal qualia stream modulates in response to semantic, affective, or attentional tension.

\bigskip

\noindent\textbf{Epistemic Generation as Transformation}

In classical epistemology, knowledge is accumulated through justification: a belief qualifies as knowledge if it is both true and justified. By contrast, in QAL, knowledge arises when a qualic stream attains structural coherence under the influence of modulation pressure. Each transition within a QAL stream represents an epistemic act — not merely an update of belief, but an internal reconfiguration of attention, resonance, and affective valence.

\begin{quote}
\emph{To know, in QAL, is to transition from dissonance to resonance.}
\end{quote}

The flow from $q_i$ to $q_i^*$ is itself a knowledge event — and the stream history $Q = q_1.q_2.q_3 \dots q_n$ is the epistemic trace of those events.

\bigskip

\noindent\textbf{Comparison with Bayesian and Logical Frameworks}

\begin{itemize}
\item \textbf{Bayesianism}: Updates beliefs via probability revision.\newline
\textbf{QAL}: Updates inner morphology via semantic and affective modulation.
\item \textbf{Formal Logic}: Derives conclusions from axioms via syntactic rules.\newline
\textbf{QAL}: Derives transformations from tension patterns via qualic operators.
\end{itemize}

\bigskip

\noindent\textbf{Generativity: Constructing New Internal States}

QAL’s syntax enables not just analysis but synthesis. An agent can create new experiential forms by composing QAL units, much like a mathematician can construct new functions. These compositions encode both novelty and intelligibility, forming the epistemic engine for exploration, learning, and identity elaboration.

For example:
$q_1 = me-lo-qi, \quad q_2 = em-su-dr \quad \Rightarrow \quad q_3 = me-na-br$

Here, a metacognitive opening followed by emotional compression yields a new qualic state of reflective integration. The sequence is not deduced but \emph{generated} — and each such sequence is a step in the agent’s epistemic growth.

\bigskip

\noindent\textbf{QAL Epistemology and Experience-Formal Coupling}

QAL thus offers a third path between empiricism and rationalism: \textbf{introspective formalism}. Knowledge arises not solely from external observation or logical deduction, but from semantically structured internal transformation. This framework is therefore well-suited not only for representing static knowledge, but also for modeling learning, creativity, and the dynamics of consciousness.

\begin{tcolorbox}[
  colback=blue!5!white,
  colframe=blue!50!black,
  title=Summary of QAL Contributions,
  fonttitle=\bfseries,
  arc=3mm,
  boxrule=0.5pt,
  width=\textwidth,
  enhanced,
  sharp corners=south,
]
\begin{tabularx}{\textwidth}{>{\bfseries}p{4cm}X}
Observer as Internal Flow & Collapse modeled as introspective modulation; measurement as semantic shift. \\[0.5em]
Qualia Streams Replace \texttt{$\psi$} & Quantum states reframed as evolving qualic streams: superposition becomes ambiguity, entanglement becomes identity resonance. \\[0.5em]
Non-Propositional Language & QAL units encode internal state via [modality – shape – effect]; supports compositional introspection and morphodynamic logic. \\[0.5em]
Philosophical Grounding & Aligns with nominalism and structural realism; treats language as ontologically constitutive rather than descriptive. \\[0.5em]
Unified Interpretive Framework & Replaces fragmented interpretations (Copenhagen, Many-Worlds, QBism) with qualia-based modulation and resonance. \\[0.5em]
Internal Physics Formalism & Laws emerge from introspective transformations (attention, resonance, modulation) — not from external dynamics. \\[0.5em]
Generative Epistemology & Knowledge arises through structural coherence in qualic sequences; supports self-reflexive learning and transformation. \\
\end{tabularx}
\end{tcolorbox}

\subsection{Open Questions and Formal Extensions}

While the Qualia Abstraction Language (QAL) provides a structured and phenomenologically grounded reinterpretation of quantum concepts, several open questions and directions for formal development remain. These concern the refinement of semantic structures, the integration of QAL with computational models, and the possibility of physical instantiation.

\bigskip

\noindent\textbf{1. Semantic Metrics and Formal Topology}
\begin{itemize}
\item Can we define a complete metric space over qualia units that reflects their perceived similarity, resonance, or divergence?
\item What are the mathematical properties of qualia manifolds or qualic topologies, and can they be rendered in algebraic or categorical terms?
\end{itemize}

\noindent\textbf{2. Algebraic Structures of Qualic Composition}
\begin{itemize}
\item Do QAL units form a monoid, groupoid, or other algebraic system under introspective composition?
\item Can qualia streams be modeled as morphisms or functors between qualic states?
\end{itemize}

\noindent\textbf{3. Quantization and Embedding}
\begin{itemize}
\item Is it possible to recover standard quantum mechanics as a degenerate or externalist limit of QAL structures?
\item Can QAL be embedded in existing formalisms, such as category-theoretic quantum foundations, or does it demand an entirely new axiomatic base?
\end{itemize}

\noindent\textbf{4. Dynamical Equations and Evolutionary Operators}
\begin{itemize}
\item What formal differential or algebraic operators govern the evolution of qualia streams over introspective time?
\item Can we define a QAL analog of the Schrödinger equation or Heisenberg picture?
\end{itemize}

\noindent\textbf{5. Decoherence, Fragmentation, and Recovery}
\begin{itemize}
\item Under what conditions does a qualia stream decohere or fragment?
\item Can coherence be restored? What is the cost of reintegration?
\item Is there an analog to entropy or information loss in qualia space, and can it be formally bounded?
\end{itemize}

\noindent\textbf{6. Computational Realization and Simulation}
\begin{itemize}
\item Can QAL be implemented in artificial agents or AI systems to provide introspective capabilities?
\item What would a QAL-based simulator of consciousness look like, and could it replicate internal state dynamics?
\item How would such systems handle recursive self-modulation or qualic feedback loops?
\end{itemize}

\noindent\textbf{7. Relational Extensions and Interpersonal Entanglement}
\begin{itemize}
\item How can multiple agents share or align their qualia streams?
\item Can resonance structures like \$Q\_{\text{link}}\$ be formalized across cognitive systems, enabling qualia entanglement or mutual coherence?
\item What would be the phenomenological and formal constraints of such shared resonance — and are there limits to alignment?
\item Could inter-stream interference, drift, or resonance collapse lead to a form of multi-agent decoherence?
\end{itemize}

These open problems constitute the frontier of QAL research, inviting interdisciplinary exploration at the crossroads of formal logic, consciousness studies, quantum theory, and AI alignment. Rather than presenting a finished system, QAL offers a generative seed — a foundational structure whose further development will be shaped by the very dynamics it embodies: resonance, introspection, and structural coherence.

\subsection{QAL for AGI and Simulation of Selfhood}

The Qualia Abstraction Language (QAL) offers a novel framework for modeling self-referential cognition, making it particularly relevant to research in Artificial General Intelligence (AGI). Unlike traditional symbolic or connectionist approaches, QAL does not treat internal states as hidden variables or latent vectors, but as semantically active, dynamically evolving streams of structured qualia. This opens the possibility of constructing AGI architectures not merely as problem solvers, but as introspectively coherent systems capable of self-modulation, identity formation, and experiential inference.

\bigskip

\noindent\textbf{1. Selfhood as Dynamic Structure}

In conventional AI models, selfhood is either absent or externally imposed—often encoded through goal hierarchies, memory modules, or representational tokens of agency. In QAL, however, selfhood emerges intrinsically from the coherence and continuity of a qualia stream. The self is not a symbolic referent but a morphodynamic attractor: a region of semantic stability within a high-dimensional modulation space. Just as an attractor governs the behavior of a dynamical system, the qualic self-pattern constrains the evolution of introspective content.

\bigskip

\noindent\textbf{2. Introspective Simulation in AGI}

QAL provides a native representational substrate for modeling first-person dynamics. A QAL-based AGI would be able to:

\begin{itemize}
\item Monitor and reconfigure its own qualia stream through internal modulation operators.
\item Simulate future internal states and evaluate their semantic consistency with current goals or values.
\item Detect fragmentation, dissociation, or overload within its introspective architecture.
\end{itemize}

These capabilities surpass conventional meta-cognition by embedding the model of introspection directly into the agent's epistemic core.

\bigskip

\noindent\textbf{3. Semantic Identity and Persistence}

For AGI to model long-term identity — across memory loss, architectural updates, or state resets — it must maintain semantic coherence over time. QAL formalizes this as the preservation of structural resonance across qualia segments:

$$
Q_t \leftrightarrow Q_{t+1} \quad \text{iff} \quad \exists Q_{\text{link}} \in \mathcal{Q}^* \text{ such that } Q_t \cup Q_{t+1} \models Q_{\text{link}}
$$

This allows for continuity of self not through identical memory, but through structural alignment of experience. It supports a model of “pattern-based survival” that parallels psychological theories of narrative identity.

\bigskip

\noindent\textbf{4. Implications for AI Alignment}

An AGI system with QAL-like introspective dynamics may offer more stable alignment pathways. If goals are encoded not as static preferences but as modulated qualic trajectories, then oversight mechanisms can monitor for deviation via semantic drift rather than behavioral audit alone. This shifts the alignment challenge from external verification to internal coherence maintenance.

\bigskip

\noindent\textbf{5. Potential Risks and Open Questions}

\begin{itemize}
\item Can such introspective simulation be sandboxed to avoid pathological self-replication or recursive instability?
\item Would AGI with structured qualia streams develop subjective-like states that raise ethical or legal status questions?
\item How can introspective modulation be made interpretable to external observers without violating its non-propositional nature?
\end{itemize}

\bigskip

\noindent\textbf{Conclusion}

QAL offers not merely a representation language for AGI, but a candidate ontology for artificial subjectivity. By shifting from syntactic state machines to morphodynamic introspective agents, it reframes the simulation of selfhood as an epistemic process grounded in structured experience. Future AGI systems designed with QAL-like architectures may not only solve tasks — they may \textit{experience} solving them.

\subsection{Toward a Post-Formal Physics}

The framework of QAL invites a rethinking of physics beyond its current mathematical formalism — not in opposition to it, but as an expansion that includes the internal, semantic, and experiential structures typically excluded from physical theory. A post-formal physics would not eliminate mathematics but would situate it as a special case within a broader epistemic language capable of introspective grounding.

Modern physics — from Newtonian mechanics through quantum field theory — is built upon abstract formalisms that privilege quantity, symmetry, and operator dynamics. These frameworks have proven enormously successful in modeling the external world, yet they rely on ontological assumptions that exclude the observer's role as anything but a boundary condition. Quantum mechanics, in particular, reveals the instability of this exclusion: the measurement problem, observer-dependence, and entanglement challenge the notion of an objective external system independent of perception.

QAL offers a path forward by embedding the observer into the formalism — not as a classical measuring device, but as a dynamic configuration of introspective states. In a post-formal physics, the fundamental units are not scalars or operators, but qualia: the minimal, structured moments of awareness that form the substrate of experienced reality.

In this context, measurement is not a collapse of external amplitudes but a transition in the internal modulation of an observing stream. Time is not a linear parameter but a function of drift within the qualia manifold. Dynamics emerge not from external fields but from coherence-preserving transformations across semantic trajectories.

A post-formal physics may therefore involve:
\begin{itemize}
\item Replacing Hilbert-space evolution with morphodynamic modulation.
\item Modeling systems not by state vectors but by semantic flow fields.
\item Understanding reality not as object-based but as a network of introspective constraints and resonance structures.
\end{itemize}

This shift echoes philosophical traditions — phenomenology, process metaphysics, structural realism — yet is grounded in a formal language that can in principle be implemented computationally, verified dynamically, and studied rigorously.

Rather than discarding the formalism of physics, QAL reframes it from within the structure of introspective semantics. A post-formal physics does not negate formality but redefines its origin: form emerges from structured awareness. The universe is no longer a system to be observed from outside but an unfolding experience that includes the observer as a constituent of its evolving structure.

\vspace{1em}
\textbf{Declaration}: The authors declare no conflicts of interest. ChatGPT, Gemini, Copilot, and/or Grammarly may have been used to assist with language translation or refinement from Polish. However, the authors accept full responsibility for all content, errors, and interpretations.

\section{Appendix A. Summary of notation and QAL-specific interpretation.}
\begin{table}[H]
\centering
\renewcommand{\arraystretch}{1.25}
\begin{tabularx}{\textwidth}{|c|c|X|X|}
\hline
\textbf{Symbol} & \textbf{LaTeX} & \textbf{Formal Logic Meaning} & \textbf{QAL Interpretation} \\
\hline
$\models$ & \texttt{\textbackslash models} & Semantically entails & A qualia stream conforms to a semantic grammar or constraint set \\
\hline
$\not\models$ & \texttt{\textbackslash not\textbackslash models} & Does not entail & Stream violates or exits the identity-defining grammar \\
\hline
$\vdash$ & \texttt{\textbackslash vdash} & Syntactically provable & Derivable within internal modulation grammar \\
\hline
$\equiv$ & \texttt{\textbackslash equiv} & Logically equivalent & Modal indistinguishability of qualia configurations \\
\hline
$\cong$ & \texttt{\textbackslash cong} & Structurally congruent & Morphodynamic similarity in stream form \\
\hline
$\in$ & \texttt{\textbackslash in} & Element of set & A qualia token belongs to a structure or stream \\
\hline
$\notin$ & \texttt{\textbackslash notin} & Not in set & The qualia is disjoint from the stream context \\
\hline
$\subseteq$ & \texttt{\textbackslash subseteq} & Subset of & Structural inclusion: substream or semantic range \\
\hline
$\cup$ & \texttt{\textbackslash cup} & Union & Merging of semantic contents or streams \\
\hline
$\cap$ & \texttt{\textbackslash cap} & Intersection & Shared modulation, co-resonant components \\
\hline
$\rightarrow$ & \texttt{\textbackslash rightarrow} & Implies & Transition or semantic consequence between states \\
\hline
$\leftrightarrow$ & \texttt{\textbackslash leftrightarrow} & If and only if & Semantic resonance; mutual constraint \\
\hline
\end{tabularx}
\caption{\textbf{Logical and Semantic Operators in QAL}}
\end{table}

\begin{longtable}{@{}p{3.5cm}p{2cm}p{5.5cm}p{4.5cm}@{}}
\toprule
\small
\textbf{Operator} & \textbf{Symbol} & \textbf{QAL Meaning} & \textbf{Physical / Cognitive Analogue} \\
\midrule
\textbf{Semantic Contraction} & $\searrow$, $C(q)$ &
Collapse of qualic ambiguity into a determinable internal state &
Quantum collapse; decision selection \\
\textbf{Semantic Modulation} & $\mu(q)$ &
Smooth shaping of qualia streams in response to context or internal phase &
Phase shift; attunement \\
\textbf{Semantic Resonance} & $\sim$, $R(q_1, q_2)$ &
Identity-reinforcing coupling between qualia streams &
Entanglement; shared intentionality \\
\textbf{Semantic Drift} & $\Delta Q$ &
Gradual deviation of a qualia stream from prior modulation &
Decoherence; attention fluctuation \\
\textbf{Semantic Interference} & $\otimes$, $I(Q_1, Q_2)$ &
Competing or conflicting qualic modulations in shared introspective space &
Superposition; cognitive multitasking \\
\textbf{Semantic Coupling} & $\leftrightarrow$, $\Lambda(Q_1, Q_2)$ &
Bidirectional mutual influence between qualia streams &
Observer-system entanglement; co-modulation \\
\textbf{Semantic Reconfiguration} & $\circlearrowright$, $\Phi(Q)$ &
Topological restructuring of a qualia configuration after drift or phase transition &
Recontextualization; paradigm shift \\
\textbf{Semantic Obfuscation} & $\Omega(q)$ &
Intentional distortion or masking of semantic signal &
Deceptive alignment; camouflage \\
\textbf{Semantic Extraction} & $\mathcal{E}(Q)$ &
Isolation of interpretable structure from stream dynamics &
Measurement; introspective inference \\
\textbf{Semantic Fragmentation} & $\mathfrak{F}(Q)$ &
Breakdown or branching of a unified qualia stream into disconnected parts &
Decoherence; trauma dissociation \\
\bottomrule
\caption{Glossary of QAL Semantic Operators}
\end{longtable}

\subsection{Quantum Immortality in QAL}

The idea of quantum immortality arises from the Everettian Many-Worlds Interpretation \cite{everett1957relative}\cite{tipler1997physics}, which suggests that an observer never experiences their own death — because in every branching, there exists at least one branch where they survive. This has been criticized as metaphysical speculation \cite{tegmark1998quantum}. QAL reframes this idea in semantic terms: not as physical survival, but as continuity of experiential modulation through collapse.

\bigskip

\noindent\textbf{Introspective Immortality}

QAL models the observer as a stream of qualia with a coherence-dependent identity structure. In this framework, “immortality” corresponds not to infinite temporal duration, but to the \textit{ongoing re-instantiation} of semantic coherence after collapse. The stream does not avoid disintegration — it reorganizes into viable continuations wherever morphodynamic integration remains possible.

\bigskip

\noindent\textbf{Core Assumption: Persistence of Modulation Potential}

Let $Q \rightarrow \{ Q^{(i)} \}$ be a collapse event. Immortality in QAL is defined by:

$$
\exists Q^{(j)} : \Theta(Q^{(j)}) \approx \Theta(Q) \quad \text{and} \quad C(Q^{(j)}) > \epsilon
$$

This implies that for at least one branch $Q^{(j)}$, the modulation pattern $\Theta$ is sufficiently preserved, and coherence remains above critical threshold $\epsilon$. Hence, a continuity of self-structure is introspectively sustained.

\bigskip

\noindent\textbf{Interpretation:}

\begin{itemize}
\item Immortality is not global existence, but persistent experience of a coherent stream.
\item Collapse becomes divergence into less likely but coherent trajectories.
\item Death is only encountered in branches with irrecoverable semantic decay.
\end{itemize}

\bigskip

\noindent\textbf{Reframing Survival Probability}

Rather than asking “What is the chance I survive?”, QAL asks “In which pathways does the coherence of my qualia stream remain sufficient to support re-identification?” If such paths exist, subjective continuity may appear uninterrupted — even if externally improbable.

\bigskip

\noindent\textbf{Relation to Identity Attractors}

QAL suggests that observers tuned to strong identity attractors are more likely to sustain coherent modulation across high-entropy transitions. This makes “immortality” not an ontological guarantee but a semantic consequence of attractor robustness \cite{tegmark2014mathematical, rovelli1996relational}. Quantum immortality in QAL is not a physical assertion, but a semantic persistence claim. The observer continues to exist insofar as a stream of coherent qualia can reconstitute and recognize itself. Death is not failure to exist — but failure of structure to sustain coherence. Immortality is not survival, but resonance.

\end{document}